\documentclass[a4paper,11pt]{article}
\usepackage[dvipdfmx]{graphicx}
\usepackage{amsmath}        
\usepackage{amssymb}        
\usepackage{cite}
\usepackage{cancel}
\usepackage{fancybox}
\usepackage{color}
\usepackage{url}
\usepackage{ulem}
\usepackage{pifont}
\usepackage{fancyhdr}
\usepackage{longtable}
\usepackage{diagbox}
\usepackage{lscape}
\usepackage{colortbl}
\usepackage{multicol}
\makeatletter
 
 \@addtoreset{equation}{section}
\makeatother

\def\bfR{{\boldsymbol R}}
\def\bfr{{\boldsymbol r}}
\def\bfN{{\boldsymbol N}}

\def\nn{\nonumber\\}
\def\tr{\mathop{\text{tr}}}

\def\slash#1{{\ooalign{\hfil/\hfil\crcr$#1$}}}
\def\half{\frac12}
\def\VEV#1{\langle{#1}\rangle}

\def\sqr#1#2{{\vcenter{\hrule height.#2pt
      \hbox{\vrule width.#2pt height#1pt \kern#1pt
          \vrule width.#2pt}
      \hrule height.#2pt}}}

\def\ket#1{\left|{#1}\right\rangle}
\def\USp{U\hspace{-0.125em}Sp}

\def\bfn{{\bf n}}
\def\bftwon{{\bf 2}\bfn}
\def\calP{{\cal P}}
\def\hatN{{\hat N}}
\def\bfGam{{\boldsymbol \Gamma}}

\newdimen\Tdim
\newdimen\Ddim
\def\Tspan#1#2#3{{\setbox0=\hbox{$#3$}%
\Tdim\ht0\advance\Tdim\dp0\advance\Tdim#1\advance\Tdim#2\Ddim\dp0\advance\Ddim#2\rule[-\Ddim]{0pt}{\Tdim}\box0}}

\makeatletter
 
 \@addtoreset{equation}{section}
\makeatother

\begin{document}
\thispagestyle{fancy}

\title{Dynamical Breaking to Special or Regular Subgroups\\ 
in $SO(N)$ Nambu--Jona-Lasinio Model}

\author{Taichiro Kugo\footnote{Electronic address: kugo@yukawa.kyoto-u.ac.jp}
and
Naoki Yamatsu\footnote{Electronic address: yamatsu.naoki@phys.kyushu-u.ac.jp}
\\
{\it\small
\begin{tabular}{c}
Yukawa Institute for Theoretical Physics, Kyoto University, Kyoto 606-8502, Japan${}^\ast$\\
Department of Physics, Kyushu University, Fukuoka 819-0395,
Japan$^\dag$
\end{tabular}
}
}
\date{\today}

\maketitle

\thispagestyle{fancy}

\begin{abstract}
 It is recently shown that in 4D $SU(N)$ Nambu--Jona-Lasinio (NJL) type
 models, the $SU(N)$ symmetry breaking into 
 its special subgroups is not special but much more common than that into the 
 regular subgroups, where the 
 fermions belong to complex representations of $SU(N)$.
 We perform the same analysis for $SO(N)$ NJL model for various 
 $N$ with fermions belonging to an irreducible spinor representation of 
 $SO(N)$. 
 We find 
 that the symmetry breaking into special or regular subgroups 
 has some correlation with the type of fermion representations; 
 i.e., complex, real, pseudo-real representations.
\end{abstract}

\section{Introduction}

Symmetries and their
breaking\cite{Nambu:1961tp,Nambu:1961fr,Goldstone:1961eq} play a crucial 
role in constructing unified theories beyond the 
Standard Model (SM) as well as the SM.
There have been already known several symmetry breaking mechanisms
in quantum field theories; e.g.,
the Higgs mechanism\cite{Higgs:1964pj,Englert:1964et,Guralnik:1964eu}, 
the dynamical symmetry 
breaking\cite{Nambu:1961tp,Nambu:1961fr,Schwinger:1962tn,Maskawa:1974vs,Maskawa:1975hx,Fukuda:1976zb,Weinberg1976,Susskind:1978ms,Raby:1979my,Dimopoulos:1979es,Farhi:1980xs,Peskin:1980gc,Miransky:1988xi,Miransky:1989ds,Kugo:1994qr,Kugo:2019isl},
the Hosotani
mechanism\cite{Hosotani:1983xw,Hosotani:1988bm,Hatanaka:1998yp}, 
magnetic flux\cite{vonGersdorff:2007uz,Abe:2008fi}
and orbifold breaking\cite{Dienes:1996yh,Kawamura:1999nj}.

Recently, in Ref.~\cite{Kugo:2019isl}, present authors 
have examined
dynamical symmetry breaking pattern in 4D $SU(N)$ Nambu--Jona-Lasinio
(NJL) type models in which the fermion belongs to an irreducible
representation of $SU(N)$ $\bfn$ or ${\bf \bfn(\bfn-1)/2}$.
The potential analysis has shown that for almost all cases at the potential 
minimum, the $SU(N)$ group symmetry is broken to its
{\it special} subgroups such as $SO(N)$ or $\USp(N)$ when symmetry
breaking occurs. (Note that special subgroups may be referred to
as non-regular or irregular subgroups depending on literature.
For more
information about Lie subgroups, see, e.g., 
Refs.~\cite{Dynkin:1957um,Dynkin:1957ek,Cahn:1985wk,Slansky:1981yr,Yamatsu:2015gut}.) 
Also, in Ref.~\cite{Kugo:1994qr}, one of the author (T.K.) and J.~Sato
have performed the same analysis in a 4D $E_6$ NJL type model  
in which the fermion belongs to an irreducible representation {\bf 27} of
$E_6$. The potential analysis showed that for all cases at
the potential minimum the $E_6$ group symmetry is broken to its special
subgroups such as $SU(3)$, $\USp(8)$, $G_2$, $F_4$ when symmetry breaking
occurs. 
These results clearly show that the symmetry breaking into special
subgroups is not special at least in 4D NJL type models.

One might think that the above results may highly depend on 
a particular feature of the NJL type models, that is, 
fermion pair condensation in a specific type of effective
theories, so the symmetry breaking into special subgroups may 
still be special, e.g., in Higgs 
mechanism\cite{Higgs:1964pj,Englert:1964et,Guralnik:1964eu}.
In Ref.~\cite{Li:1973mq}, however, L.-F.~Li investigated the
symmetry breaking patterns in gauge theories with Higgs scalars
in several representations.
The analysis showed that $SU(n)$ symmetry is broken to its regular
subgroups for the cases of Higgs field in an 
$SU(n)$ fundamental and adjoint
representation, while $SU(n)$ symmetry is broken to its regular or
special subgroups for the cases of Higgs field 
in an $SU(n)$ rank-2 symmetric or
anti-symmetric tensor representation. 
Therefore, the symmetry breaking into special subgroups is not so
special even in the Higgs models. 

One of the attractive ideas in constructing theories beyond 
the SM is the grand unification theories (GUTs) proposed 
in Ref.~\cite{Georgi:1974sy}.
To realize the SM, which is a 4D chiral gauge theory, as a low energy
effective theory, GUTs must be chiral gauge
theories from the viewpoint of 4D effective theories. In other words,
in 4D frameworks, 4D GUTs must contain at least a Weyl fermion in a
complex representation of the GUT gauge group $G_{\rm GUT}$. Thus,
the candidates of 4D GUT gauge group $G_{\rm GUT}$ may allow the
existence of complex representations.
(For the type of representations of Lie groups, see e.g.,
Refs.~\cite{McKay:1981,Slansky:1981yr,Yamatsu:2015gut}.)
The condition suggests that $G_{\rm GUT}$ may be
$SU(N) (N\geq5)$, $SO(4n+2)$ $(n\geq2)$, or $E_6$.
There are many proposals for 4D GUTs 
\cite{Georgi:1974sy,Fritzsch:1974nn,Gursey:1975ki,Inoue:1977qd,Ida:1980ea,Fujimoto:1981bv,Yamatsu:2017sgu,Yamatsu:2018fsg}
and 5D or higher dimensional GUTs
\cite{Kawamura:2000ev,Kawamura:2000ir,Burdman:2002se,Kim:2002im,Lim:2007jv,Fukuyama:2008pw,Kojima:2011ad,Hosotani:2015hoa,Yamatsu:2015rge,Furui:2016owe,Kojima:2016fvv,Kojima:2017qbt,Hosotani:2017ghg,Hosotani:2017edv,Yamatsu:2017sgu,Yamatsu:2017ssg,Yamatsu:2018fsg}.
Note that in higher dimensional theories on e.g., orbifold spaces 
unconventional GUT groups such as
$SO(11)$\cite{Hosotani:2015hoa,Yamatsu:2015rge,Furui:2016owe,Hosotani:2017ghg,Hosotani:2017edv}
and $SO(32)$\cite{Yamatsu:2017ssg} are allowed because zero modes of some
higher dimensional fields with appropriate boundary conditions can be 4D
Weyl fermions in complex representations of their Lie subgroups. 

From the usage in GUT model buildings in 4D framework, we 
are primarily motivated in studying 4D NJL type models 
in which the fermion matter
belongs to complex representations of $SU(N) (N\geq5)$, $SO(4n+2)$ $(n\geq2)$, 
and $E_6$. We have already known some examples for $SU(N)$
and $E_6$ cases in Ref.~\cite{Kugo:2019isl,Kugo:1994qr}, so if we
investigate $SO(4n+2) (n\geq2)$ cases, our primary purpose 
will be completed at least for 4D cases. As is well-known in e.g.,
Refs.~\cite{McKay:1981,Slansky:1981yr,Yamatsu:2015gut}, 
the irreducible $SO(N)$ spinor representations are
classified into three types depending on $N$ (mod 8); i.e.,
complex representations for
$N=8\ell\pm2\ (\ell\geq1))$;
real representations for
$N=8\ell, 8\ell\pm1 (\ell\geq1)$;
pseudo-real representations for
$N=8\ell\pm3, 8\ell+4 (\ell\geq1)$,
which are summarized in Table~\ref{Table:SON-Spinor-Representations}.
Some higher dimensional models, e.g., 
5 and 6 dimensional $SO(11)$ gauge-Higgs GUTs
\cite{Hosotani:2015hoa,Hosotani:2017edv}
contain the fermions in the spinor representation of $SO(11)$ {\bf 32},
which is a pseudo-real representation.
So, we are also interested in symmetry breaking pattern 
caused by the pair condensation of fermions not only in complex
representations but also in real and pseudo-real representations.
Through the investigation of 4D $SO(N)$ NJL type models in
which the fermion matter belongs to the spinor representation,
we find that the symmetry breaking into special or regular subgroups 
has certain correlation with the type of representations of 
the fermion. 
(Note that even Ref.~\cite{Li:1973mq} has not investigated
the Higgs potential models whose scalar field belongs to the 
$SO(N)$ spinor representation except for $N=5$ case because the dimension 
of the $SO(N)$ spinor representation, $2^{[(N-1)/2]}$, rapidly becomes 
too large for large $N$.)

\begin{table}[t]
\begin{center}
\begin{tabular}{cclc}
\hline
$\begin{tabular}{c}
 $N$ (mod.~$8$)\\
\end{tabular}$
 &
$\begin{tabular}{c}
 Spinor rep.
\end{tabular}$
&
dimension
&
$\begin{tabular}{c}
 Spinor type\\
 R/PR/C\\
\end{tabular}$
\\\hline
0&${\bfr},\ {\bfr}'$   & \quad $2^{\frac{N}2-1}$ & R\\
1&${\bfr}$             & \quad $2^{\frac{N-1}2}$ & R\\
2&${\bfr},\ \bar{\bfr}$& \quad $2^{\frac{N}2-1}$ & C\\
3&${\bfr}$             & \quad $2^{\frac{N-1}2}$ & PR\\
4&${\bfr},\ {\bfr}'$   & \quad $2^{\frac{N}2-1}$ & PR\\
5&${\bfr}$             & \quad $2^{\frac{N-1}2}$ & PR\\
6&${\bfr},\ \bar{\bfr}$& \quad $2^{\frac{N}2-1}$ & C\\
7&${\bfr}$             & \quad $2^{\frac{N-1}2}$ & R\\
\hline
\end{tabular}
\caption{$SO(N)$ spinor representations: R, PR, C stand for real,
 pseudo-real, and complex representations. For even $N$, 
 there are two irreducible spinors, denoted here as  
 ${\bfr}$ and ${\bfr}'$(or $\bar{\bfr}$), with opposite `chirality'. }
\label{Table:SON-Spinor-Representations}
\end{center}
\end{table}

In this paper, we analyze the $SO(N)$ symmetry breaking patterns in 4D
$SO(N)$ NJL type models 
with fermions of $SO(N)$ irreducible spinor representation 
which is complex for $N=4n+2$
and self-conjugate otherwise.
The main purpose of this study is 
to show that for $N=4n+2\ (n\in\mathbb{Z}_{\geq1})$, 
$SO(N)$ symmetry is 
broken to its {\it special} subgroups such as
$SO(\frac{N}{2})\times SO(\frac{N}{2})(S)$, 
$G_2(S)$ for $N=14$ and $F_4(S)$ for $N=26$. 
For $N\not=4n+2$, on the other hand, the prime number cases
$N \in\mathbb{P}_{\geq3}$ and the other composite number cases 
$N\in\mathbb{Z}_{\geq3}\backslash(\mathbb{P})$ are distinguished. 
For prime numbers 
$N \in\mathbb{P}_{\geq3}$ such as $3,5,7,11,13,\cdots$, 
$SO(N)$ symmetry is always broken to its regular subgroups such as
$SO([\frac{N+1}{2}])\times SO([\frac{N-1}{2}])(R)$.
For composite numbers $N\in\mathbb{Z}_{\geq3}\backslash(\mathbb{P})$ 
(other than $N=4n+2$) such as $4,8,9,12,16,\cdots$,
$SO(N)$ symmetry is
broken to its regular or special subgroups such as
$SO([\frac{N+1}{2}])\times SO([\frac{N-1}{2}])(R)$. (
Note that $(R)$ and $(S)$ here stand for regular and special subgroups,
respectively.)

This paper is organized as follows. 
In Sec.~\ref{Sec:SOn-spinor}, we quickly review the $SO(N)$ NJL
models and examine $SO(N)$ spinor properties and 
maximal little groups of $SO(N)$ for the irreducible representations 
of the fermion bilinear composite scalar fields.
By using the knowledges in Sec.~\ref{Sec:SOn-spinor}, 
we discuss addition-type, product-type, and embedding-type subgroups
in Sec.~\ref{Sec:Additon-type-subgroups},
\ref{Sec:Product-type-subgroups}, and
\ref{Sec:Embedding-type-subgroups}, respectively.
Section~\ref{Sec:Summary-discussion} is devoted to summary and
discussion.

Note that we treat the same type of 4D NJL models and 
use the same notation as in Ref.~\cite{Kugo:2019isl}, so we will not 
repeat them in detail in this paper.
In the paper, we denote Dynkin labels such as
$(0,0,\cdots,0,2)=:(0^{n-1},2)$,
$(0,0,\cdots,0,1,0,0,0,0)=:(0^{n-5},1,0^{4})$, $\cdots$
for rank-$n$ groups, for simplicity.

\section{NJL-type model and $SO(N)$ spinor properties}
\label{Sec:SOn-spinor}

We consider the NJL-type model where the fermion $\psi_I$ 
($I=1,\cdots,d$) belongs to the irreducible 
$SO(N)$ spinor representation $\bfR$ of dimension 
$d:=2^{[\frac{N-1}{2}]}$, 
denoted by using Dynkin label as
\begin{equation}
\bfR = 
\begin{cases}
\bfr = (0^{n-1}, 1)\  \text{or}\ \bfr'(\text{or}\ \bar\bfr)
=(0^{n-2}, 1, 0) \ & 
\hbox{for even}\ N=2n \\
\bfr = (0^{n-1}, 1) & \hbox{for odd}\ N=2n+1
\end{cases}.
\end{equation}
\noindent
Since the fermion $\psi_I$, for each $I$, actually stands for 2-component 
Weyl spinor of Lorentz group, the fermion bilinear scalar
$\Phi_{IJ}\sim\psi_I\psi_K$ gives {\it symmetric} product
$(\bfR\times\bfR)_{\rm S}$ decomposed into the following 
$1+[\frac{N}{8}]$ irreducible representations $\bfR_0$ and $\bfR_p$ 
($p=1,2,\cdots,[\frac{N}8]$), and, additionary for odd $N$, 
$[\frac{N+1}{8}]$ irreducible representations $\bfR_p'
(q=1,2,\cdots,[\frac{N+1}{8}])$:
\begin{align}
\left({\bf 2}^{[\frac{N-1}{2}]}\otimes{\bf
 2}^{[\frac{N-1}{2}]}\right)_{\rm S}=
 \left\{
\begin{array}{ll}
\displaystyle
 {\bfR}_{0}
 \oplus
\bigoplus_{p=1}^{[\frac{N}{8}]}{\bfR}_{p} & \mbox{for}\ \mbox{even}\ N=2n\\
\displaystyle
 {\bfR}_{0}
 \oplus
 \bigoplus_{p=1}^{[\frac{N}{8}]}{\bfR}_{p}
 \oplus
 \bigoplus_{q=1}^{[\frac{N+1}{8}]}{\bfR}_{q}'& \mbox{for}\ \mbox{odd}\ N=2n+1\\
\end{array}
 \right., 
\end{align}
where ${\bfR}_{0}$,
${\bfR}_{p}$ and ${\bfR}_{q}'$ 
are the representations 
denoted by the following Dynkin labels and correspond to the denoted 
anti-symmetric (AS) tensors:
\begin{align}
&{\bfR}_{0}:=
\begin{cases}
(0^{n-1},2) & \text{for\ } \bfR=\bfr \\
(0^{n-2},2,0) & \text{for\ } \bfR=\bfr' \text{or\ } \bar\bfr 
\end{cases} 
\ \sim\  \psi_I (C\Gamma_{a_1\cdots a_{n}})^{IJ}\psi_J \quad 
\text{rank-$n$ AS tensor (self-dual if $N=2n$)}
,\ \ \nn
&{\bfR}_{p}:=
\begin{cases}
(0^{n-1-4p},1,0^{4p}) & \hbox{for}\ 4p\leq n-1 \\
(0^n) = {\bf 1} & \hbox{for}\ 4p=n \\
\text{non-existing} & \hbox{for}\ 4p\geq n+1 \\
\end{cases}
\ \sim\  \psi_I (C\Gamma_{a_1\cdots a_{n-4p}})^{IJ}\psi_J \quad \text{rank-$(n-4p)$ AS tensor},
 \nonumber\\
&{\bfR}_{q}':=
\begin{cases}
(0^{n-4q},1,0^{4q-1}) & \hbox{for}\ 4q\leq n \\
(0^n) = {\bf 1} & \hbox{for}\ 4q=n+1 \\
\text{non-existing} & \hbox{for}\ 4q\geq n+2 \\
\end{cases}
\ \sim\  \psi_I (C\Gamma_{a_1\cdots a_{n-4q+1}})^{IJ}\psi_J \quad \text{rank-$(n-4q+1)$ 
AS tensor},
\label{eq:RpRepr}
\end{align}
where $C$ is $SO(N)$ charge conjugation matrix and 
$\Gamma_{a_1\cdots a_r}$ are rank-$r$ AS tensor gamma matrices defined by 
\begin{equation}
\Gamma_{a_1a_2\cdots a_r}:=\Gamma_{[a_1}\Gamma_{a_2}\cdots\Gamma_{a_r]}=
\begin{cases}
\Gamma_{a_1}\Gamma_{a_2}\cdots\Gamma_{a_r} &\text{if $a_1, a_2,\cdots a_r$ are all 
different} \\
0 & \text{otherwise}
\end{cases}\ 
\end{equation}
in terms of the $SO(N)$ gamma matrices $\Gamma_a$ ($a=1,2,\cdots,N$) given 
explicitly later below.
(For  $SO(N)$ tensor products, see e.g., Ref.~\cite{Yamatsu:2015gut}.)
In the following, $\bfR_k$ 
collectively
denotes $\bfR_0$ and $\bfR_p$ for even $N$; 
$\bfR_0$, $\bfR_p$ and $\bfR_q'$ for odd $N$, respectively.

Now the Lagrangian of 
the NJL model we discuss is given by
\begin{align}
{\cal L}=& 
\bar\psi^I\,i\overline{\sigma}^\mu\partial_\mu\psi_I
+ \sum_{k} \frac14{G_{\bfR_k}} 
 \bigl(\psi_I\psi_J\bigr)_{\bfR_k}
 \bigl(\bar\psi^I\bar\psi^J\bigr)_{\overline\bfR_k},
\end{align}
where $\bigl(\psi_I\psi_J\bigr)_{\bfR_k}$ is the projection of the 
fermion bilinear scalar $\psi_I\psi_J$ into the irreducible channel $\bfR_k$, and 
$G_{\bfR_k}$ is the 4-Fermi coupling constants in that channel. 
As in the previous papers~\cite{Kugo:2019isl,Kugo:1994qr}, we 
introduce a set of irreducible auxiliary scalar fields 
$\Phi_{\bfR_{k}}{}_{\,IJ}$ standing for the composite operators 
$-(G_{\bfR_k}/2)(\psi_I\psi_J)_{G_{\bfR_k}}$. Then,  
we can equivalently rewrite this Lagrangian into the form
\begin{equation}
{\cal L}=
\bar\psi^I i\bar{\slash\partial}\psi_I
-\half\psi_I\psi_J \Phi^{\dag\,IJ}
-\half\bar\psi^I\bar\psi^J \Phi_{IJ}
-\sum_{k}
M^2_{\bfR_k}\,\tr \bigl(\Phi_{\bfR_k}^{\dag}\Phi_{\bfR_k}\bigr),
\label{eq:YukawaFormL}
\end{equation}
and obtain the effective potential 
of this system
in the leading order in $1/N$ as \cite{Kugo:2019isl} 
\begin{align}
 V^{\text{leading}}(\Phi)&=
\sum_{k}M^2_{\bfR_k}\,\tr 
\bigl(\Phi_{\bfR_k}^{\dag}\Phi_{\bfR_k}\bigr) 
 +V^{\text{1-loop}}(\Phi), \nn
 V^{\text{1-loop}}(\Phi)&=
 -  \tr_d f(\Phi^\dagger\Phi)  
 =- \tr_d f(\Phi\Phi^\dagger)  
 =-  \sum_{I=1}^d f(m^2_I),
\label{eq:effpot}
\end{align}
where $M^2_{\bfR_k}:=1/G_{\bfR_k}$, 
and $\Phi_{\bfR_k}$'s are constrained symmetric $d\times d$ 
matrices subject to non-trivial condition belonging to the irreducible
representation $\bfR_k$, so satisfying the orthogonality
$\tr (\Phi^\dagger_{\bfR_k}\Phi_{\bfR_{k'}})=0$ for $k\not=k'$, 
while 
$\Phi:=\sum_k \Phi_{\bfR_k}$ without the irreducible suffix $\bfR_k$ is 
the general (unconstrained)
symmetric  $d\times d$ matrix which appears in the Yukawa interaction terms 
in Eq.~(\ref{eq:YukawaFormL}) and hence in the 1-loop part
potential $V^{\text{1-loop}}$. 
Here $m^2_I$ are $d$ eigenvalues of the Hermitian matrix
$\Phi^\dagger\Phi$, which stand for $d$ mass-square eigenvalues of the
fermion $\psi_I$, and the function $f(m^2)$ is given by 
\begin{align}
f(m^2)
&:=
\int_{p^2_{\rm E}\leq\Lambda^2}\frac{d^4p^{\phantom 3}_{\rm E}}{(2\pi)^4} 
\ln\Bigl(\frac{p_{\rm E}^2 +m^2}{p_{\rm E}^2} \Bigr) \nn 
&=
{1\over32\pi^2}\left\{
\Lambda^4\ln\Bigl(1+\frac{m^2}{\Lambda^2}\Bigr) 
-m^4\ln\Bigl(1+\frac{\Lambda^2}{m^2}\Bigr)+m^2\Lambda^2
\right\},
\end{align}
where $\Lambda$ is UV cutoff on the Euclideanized momentum.
We also use $F(x;M^2)$ defined as
\begin{align}
F(x\,; M^2) :=& M^2 x - f(x).
\label{eq:function-F}
\end{align}

To discuss spontaneous symmetry breaking, we have to check
{\it little groups}, where a little group $H_\phi$ of a 
vector $\phi$ in a representation $\bfR$ of $G$ is defined by 
\begin{equation}
H_\phi:= \Bigl\{ g \ \Big| \ g\phi=\phi,\ g\in G\ \Bigr\}.
\end{equation}
This {\it little group} $H_\phi$ of $G$ depends not only on the
representation $\bfR$ of $\phi$ but also the vector (value) $\phi$
itself. The vector $\phi$ must be an $H_\phi$-singlet, so that a
subgroup $H$ can be a little group of $G$ for some representation $\bfR$
only when $\bfR$ contains at least one  $H$-singlet.
Following so-called Michel's conjecture
\cite{Michel:1980pc}, we assume that the potential minimum of the scalar
field in an irreducible representation $\bfR$ of $G$ is located
at one of the little group $H$ of $\bfR$.

We classify little groups $H$ of $G=SO(N)$ into three types:
``addition-type,'' ``product-type,'' and ``embedding-type'' subgroups. 
The examples of these three types of subgroups
are listed in 
Tables~\ref{tab:SO2n-maximal-subgroups-addition-type}
and \ref{tab:SO2n+1-maximal-subgroups-addition-type}, 
Table~\ref{tab:SO-maximal-subgroups-product-type}, and 
Table~\ref{tab:SO-maximal-subgroups-simple-type}, respectively, from which 
the meaning of the name for these three types will be clear.
An addition-type subgroup is usually a regular subgroup except 
for the cases $N=4\ell+2$ for which $n:= N/2$ is odd and 
the subgroup $H=SO(n+4p)\times SO(n-4p)$ has lower rank than $G=SO(N=2n)$.
Any embedding-type subgroup is a special subgroup.

In this paper, we follow the notation and convention of 
Ref.~\cite{Yamatsu:2015gut} 
for the Lie algebra. Specifically, 
we normalize
the Dynkin index of the $SO(n)$ 
defining representation $\bfn$ as $T(\bfn)=1$,
the Dynkin index of the $SU(n)$ 
defining representation $\bfn$ as
$T(\bfn)=1/2$, and
the Dynkin index of $\USp(2n)$ $\bftwon$ representation as
$T(\bftwon)=1/2$.
Note that $T({\bf 3})=1$ for $SO(3)$, while $T({\bf 3})=2$ for $SU(2)$.

\begin{table}[htb]
\begin{center}
\begin{tabular}{llcc} \hline
& \hbox{Maximal little group $H$ of $SO(N=2n)$} & {\hbox{$H$-singlet in}}
 & \hbox{$n$} \\ \hline 
 $p$:&(Regular/Special)\hbox{$SO(n+4p)\times SO(n-4p)$ case} \phantom{\Big|}&
	 ${\bfR_{p}}$&(even/odd) \\[1ex]
\hline
\end{tabular}
\end{center}
\caption{
 ``Addition-type'' little groups $H$ of $SO(N=2n)$ for 
 irreducible representation ${\bfR_{p}}$ scalars possessing
 $H$-singlet are listed, where $p=0,1,2,\cdots,\left[\frac{N}{8}\right]$.
(Regular/Special) in front of each little group name 
shows that 
it is a regular or special subgroup of $SO(N=2n)$ 
depending on whether $n$ is even or odd, respectively. Note that the rank 
of the subgroup is lowered than that of 
$SO(2n)$ when $n$ is odd.} 
\label{tab:SO2n-maximal-subgroups-addition-type}
\end{table}

\begin{table}[htb]
\begin{center}
\begin{tabular}{llc} \hline
& \hbox{Maximal little group $H$ of $SO(N=2n+1)$} & {\hbox{$H$-singlet in}}  \\ \hline
 $p$)&(Regular) \hbox{$SO(n+4p+1)\times SO(n-4p)$ case} \phantom{\Big|}&
	 ${\bfR_{p}}$ \\[1ex]
 $q'$)&(Regular) \hbox{$SO(n+4q)\times SO(n-4q-1)$ case} \phantom{\Big|}&
	 ${\bfR_{q}'}$ \\[1ex]
\hline
\end{tabular}
\end{center}
\caption{
 ``Addition-type'' little groups $H$ of $SO(N=2n+1)$ for 
  irreducible representation ${\bfR_{p}}$ and ${\bfR_{q}'}$ 
  scalars possessing $H$-singlet are listed, where 
 $p=0,1,2,\cdots,\left[\frac{N}{8}\right]$ and
 $q=1,2,\cdots,\left[\frac{N+1}{8}\right]$.
(Regular) in front of each little group name shows that they 
are all regular subgroups of $SO(N=2n+1)$.
  } 
\label{tab:SO2n+1-maximal-subgroups-addition-type}
\end{table}

\begin{table}[htb]
\begin{center}
\begin{tabular}{ll} \hline
& \hbox{Maximal little group $H$ of $SO(N)$}   \\ \hline
 $N=mn\ (m,n\geq3;m,n\not=4)$:&
 (Special) \hbox{$SO(m)\times SO(n)$ case} \phantom{\Big|} \\[1ex]
 $N=4mn\ (m,n\geq1\ \mbox{except}\ m=n=1)$:
 & (Special) \hbox{$\USp(2m)\times\USp(2n)$ case} \phantom{\Big|} \\[1ex]
 $N=2n\ (n\geq2)$:&
 (Regular) \hbox{$SU(n)\times U(1)$ case} \phantom{\Big|} \\[1ex]
\hline
\end{tabular}
\end{center}
\caption{``Product-type'' little groups $H$ of $SO(N)$ are
 listed. 
 $SO(N)$ irreducible representation 
 ${\bfR_0}$, ${\bfR_p}$, and ${\bfR_q'}$ scalars
 possessing $H$-singlet up to $N=26$ are listed in
 Table~\ref{Table:Representations-SOn-Spinor-Tensor} 
 (Note there the isomorphism $\USp(2)\simeq SO(3)\simeq SU(2)$, 
 $SO(5)\simeq \USp(4)$ and $U(1)\simeq SO(2)$). 
 (Special) and (Regular) in front of each little group name
 show that they are all special or regular subgroups 
 of $SO(N)$ with $N$ indicated.
 } 
\label{tab:SO-maximal-subgroups-product-type}
\end{table}

\begin{table}[htb]
\begin{center}
\begin{tabular}{ll} \hline
& \hbox{Maximal little group $H$ of $SO(N)$} \\ \hline
 $N=7$)& (Special) \hbox{$G_2$ case} \phantom{\Big|}\\[1ex]
 $N=14$)& (Special) \hbox{$G_2$ case} \phantom{\Big|}\\[1ex]
 $N=15$)& (Special) \hbox{$SO(6)$ case} \phantom{\Big|}\\[1ex]
 $N=16$)& (Special) \hbox{$SO(9)$ case} \phantom{\Big|}\\[1ex]
 $N=20$)& (Special) \hbox{$SO(6)$ case} \phantom{\Big|}\\[1ex]
 $N=21$)& (Special) \hbox{$\USp(6)$ case} \phantom{\Big|}\\[1ex]
 $N=21$)& (Special) \hbox{$SO(7)$ case} \phantom{\Big|}\\[1ex]
 $N=24$)& (Special) \hbox{$SU(5)$ case} \phantom{\Big|}\\[1ex]
 $N=26$)& (Special) \hbox{$F_4$ case} \phantom{\Big|}\\[1ex]
\hline
\end{tabular}
\end{center}
\caption{
 ``Embedding-type'' little groups $H$ of $SO(N)$ 
 where the $SO(N)$ defining representation $\bfN$ is identified with an 
 irreducible representation of $H$. Here such  
 Embedding-type little groups $H$ of $SO(N)$ are listed 
 up to $N=26$ (except $H=SU(2)$ cases), and which irreducible scalar 
 among $SO(N)$ ${\bfR_0}$, ${\bfR_p}$, and ${\bfR_q'}$ contains the  
 $H$-singlet is shown in
 Table~\ref{Table:Representations-SOn-Spinor-Tensor}.
(Special) in front of each little group name indicates that they 
are all special subgroups of $SO(N)$. } 
\label{tab:SO-maximal-subgroups-simple-type}
\end{table}

In the next section, we discuss the effective potential VEV of
addition-type, 
product-type, and embedding-type subgroups one by one.

Before starting the discussion, let us here recapitulate the  
basic properties of spinor representation briefly.

The $SO(N=2n)$ or $SO(N=2n+1)$ spinor representation can be described
by $2n+1$ Hermitian $2^n\times2^n$ gamma matrices 
$\Gamma_a $ satisfying
\begin{align}
\left\{\Gamma_a,\Gamma_b\right\}=2\delta_{ab}\, {\bf1}_{2^n}, 
\qquad  \Gamma_a^\dagger= \Gamma_a \ \ (a=1,2,\cdots,2n+1),\ \ n:=[N/2].
\end{align}
When necessary in this paper, we will use the following expression 
for the $SO(N)$ gamma matrices $\Gamma_a$ and the corresponding charge conjugation matrix $C$:
\begin{align}
 \Gamma_{1,2,3}&=\sigma_{1,2,3}\otimes\sigma_1\otimes\sigma_1\otimes\sigma_1\cdots\otimes\sigma_{1}\otimes\sigma_{1},
 \nonumber\\
 \Gamma_{4,5}&=\sigma_{0}\otimes\sigma_{2,3}\otimes\sigma_1\otimes\sigma_1\cdots\otimes\sigma_{1}\otimes\sigma_{1},
 \nonumber\\
 \Gamma_{6,7}&=\sigma_{0}\otimes\sigma_{0}\otimes\sigma_{2,3}\otimes\sigma_1\cdots\otimes\sigma_{1}\otimes\sigma_{1},
 \nonumber\\
 &\vdots
 \nonumber\\
 \Gamma_{2n-2,2n-1}&=\sigma_{0}\otimes\sigma_{0}\otimes\sigma_0\otimes\sigma_0\cdots\otimes\sigma_{2,3}\otimes\sigma_{1},
 \nonumber\\
 \Gamma_{2n,2n+1}&=\sigma_{0}\otimes\sigma_{0}\otimes\sigma_0\otimes\sigma_0\cdots\otimes\sigma_{0}\otimes\sigma_{2,3},
\label{Eq:Gamma-matrix}\\
C &= \sigma_2 \otimes \sigma_3\otimes\sigma_2\otimes\sigma_3\cdots\otimes 
\begin{cases}
\sigma_3\otimes\sigma_2 & \text{for $n$: odd} \\
\sigma_2\otimes\sigma_3 & \text{for $n$: even} \\
\end{cases},
\label{eq:C-matrix}
\end{align}
where $\sigma_j (j=1,2,3)$ are the Pauli matrices,
\begin{align}
\sigma_{1}=
\left(
\begin{array}{cc}
0&1\\
1&0\\
\end{array}
\right),\ 
\sigma_{2}=
\left(
\begin{array}{cc}
0&-i\\
i&0\\
\end{array}
\right),\
\sigma_{3}= 
\left(
\begin{array}{cc}
1&0\\
0&-1\\
\end{array}
\right). 
\end{align}
and 
$\Gamma_{2n+1}$ plays the role of `chirality' operator for $SO(2n)$ case.  
The charge conjugation matrix $C$ (\ref{eq:C-matrix}) in this representation 
satisfies
\begin{align}
& 
 C\,\Gamma_aC^{-1} = \eta\Gamma_a^{\rm T},\ \ \ C^{\rm T}= \varepsilon C,
 \nonumber\\
&C=C^\dagger=C^{-1},\qquad \hbox{with}\quad  
\eta= (-1)^n,\quad \varepsilon=(-1)^{[(n+1)/2]}\,.
\end{align} 

As we will see below, the auxiliary scalar field $\Phi_{\bfR_k}$ 
in any $SO(N)$ irreducible
representation $\bfR_k$ has an $H$-singlet of
an addition-type little group listed
in Tables~\ref{tab:SO2n-maximal-subgroups-addition-type}
and \ref{tab:SO2n+1-maximal-subgroups-addition-type}.
Also, in any breaking into an addition-type subgroup,
squared masses of the $SO(N)$ spinor fermion will be found all degenerate.
As has been discussed in Refs.~\cite{Kugo:2019isl,Kugo:1994qr},
if the fermion masses are all degenerate, the corresponding 
vacuum realizes the global minimum of the effective
potential in the leading order in NJL model, and so giving a candidate 
for the true vacuum.

In the following sections, we will 
often use the ``traceless condition'' for the mass matrix of our 
$SO(N)$ spinor fermion $\psi$. The fermion mass term is given by the VEV 
of the auxiliary scalar field $\Phi=\sum_k \Phi_{\bfR_k}$ as is seen from the 
Yukawa term 
\begin{equation}
-(1/2)\psi_I\psi_J \Phi^{\dag\,IJ} + \text{h.c.}
\end{equation}
in the Lagrangian (\ref{eq:YukawaFormL}).
This VEV is developed in 
an $H$-singlet component of one of irreducible scalar fields $\Phi_{\bfR_k}$, 
which, 
as already shown in Eq.~(\ref{eq:RpRepr}), 
corresponds to a certain rank-$r$ AS tensor $\Gamma$-matrix 
$\Gamma_{a_1a_2\cdots a_r}$, so that the fermion mass matrix takes the form
\begin{align}
\VEV{\Phi^\dagger_{\bfR_k}}= C\,M_r, \qquad 
M_r=\sum c^{a_1a_2\cdots a_r}\Gamma_{a_1a_2\cdots a_r}
\label{eq:Yukawa}
\end{align}
with charge conjugation matrix $C$ and $H$-singlet 
linear combination $M_r$ of rank-$r$ tensor gamma matrices. 
Note that $C$ should be present here since the scalar $\Phi$ is symmetric.  
Note, however, that both $\VEV{\Phi^\dagger_{\bfR_k}}= C\,M_r$ and $M_r$ realize 
the same quadratic mass matrix which actually determines the value of 
the effective potential 
(\ref{eq:effpot}):
\begin{equation}
\VEV{\Phi_{\bfR_k}}\VEV{\Phi^\dagger_{\bfR_k}}
= M_r^\dagger C^\dagger C\,M_r = M_r^\dagger M_r.
\end{equation}
Moreover, the mass matrix which becomes proportional to a unit matrix 
on each $H$-irreducible sector of the fermion $\psi$ is not $C M_r$ 
but $M_r$. This is because the $SO(N)$ invariance of the Yukawa term 
Eq.~(\ref{eq:Yukawa}) implies that their transformation law under the 
$g\in SO(N)$ transformation $\psi\rightarrow\psi'=g\psi$ are given by 
\begin{equation}
\Phi^\dagger\ \rightarrow\ \Phi'{}^\dagger= {g^{\rm T}}^{-1}\Phi^\dagger g^{-1} 
\quad \hbox{so that}\quad 
\begin{cases}
CM_r \ &\rightarrow\ C M'_r= {g^{\rm T}}^{-1}(CM_r)g^{-1}  \\
M_r \ &\rightarrow\ \ M_r'= g M_r g^{-1} 
\end{cases}
\label{eq:2.19}
\end{equation}
where use has been made of $C^{-1}{g^{\rm T}}^{-1} C = g$ which follows 
from the fact that 
the $SO(N)$ spinor rotation $g$ is written as
$g=\exp(\theta^{ab}\Gamma_{ab})$ with 
real angle $\theta^{ab}$ and rank-2 
gamma matrix $\Gamma_{ab}$, and the property of the charge conjugation matrix 
$C^{-1}\Gamma_{ab}^{\rm T}C=-\Gamma_{ab}$.
The $H$-invariance of the VEV implies that Eq.~(\ref{eq:2.19}) for $g=h\in H$ 
holds with $M_r'=M_r$. It is $M_r$ but not $C M_r$ that satisfies 
\begin{equation}
M_r h = h M_r \quad \text{for}\quad \forall h\in H 
\end{equation} 
taking the form for which Schur's lemma can be applied and tells us that, 
when the components 
of the fermion $\psi_I$ are decomposed and ordered into $H$-irreducible blocks, 
then the matrix $M_r$ becomes block diagonal and is proportional to an 
identity matrix in each $H$-irreducible block (so being actually diagonal 
matrix on this basis). 

Therefore we can call $M_r$ `fermion mass matrix', since it gives 
a constant mass eigenvalue on each $H$-irreducible sector of $\psi$ 
and its square $M_r M_r^\dagger$ gives square mass eigenvalues $m_I^2$
appearing in the effective potential. 

Now we can state our ``traceless condition'' for the fermion mass matrix $M_r$: 
Since the rank-$r$ AS tensor gamma matrices are traceless
\footnote{These equations can be understood as special ($s=0$) case 
of the general orthonormal relation among anti-symmetric 
tensor gamma matrices:
\begin{equation}
\tr[ \Gamma_{a_1a_2\cdots a_r}\Gamma_{b_1b_2\cdots b_s}] =
\begin{cases} 
0 & \text{for}\ r\not=s \\
s!\delta^{a_1}_{[b_1}\delta^{a_2}_{b_2}\cdots\delta^{a_r}_{b_s]} & \text{for}\ r=s 
\end{cases}
\end{equation}
}
\begin{align}
 \tr\left[\Gamma_{a_1a_2\cdots a_r}\right]=0,\ \
\end{align}
the fermion mass matrix $M_r$ given as their linear combination 
(\ref{eq:Yukawa}) is also traceless and hence the sum of mass eigenvalues 
should vanish. 

This conclusion is valid for $SO(N{=}2n{+}1)$ cases, but does not apply 
as it stands for $SO(N{=}2n)$ cases in which the irreducible spinor fermion $\psi$
is `Weyl fermion' possessing `chirality' $\Gamma_{2n+1}=+1$ or $-1$. 
When we use the representation of $\Gamma_{2n+1}$ given in 
Eq.~(\ref{Eq:Gamma-matrix}), 
Weyl spinors with chirality $\Gamma_{2n+1}=\pm1$ have 
only upper-half or lower-half non-vanishing components.
We refer to $\Gamma$-matrices possessing non-vanishing matrix elements 
only on diagonal (off-diagonal) blocks as 
``$\gamma_5$-diagonal ($\gamma_5$-off-diagonal)''. 
Then, assuming positive chirality for our spinor $\psi$, the fermion 
mass matrix in Eq.~(\ref{eq:Yukawa})
is now replaced by the following chiral projected one:
\begin{equation}
M_r=\sum_{a_1, a_2, \cdots, a_r=1}^{2n} c^{a_1a_2\cdots a_r}\Gamma_{a_1a_2\cdots a_r}\left(\frac{\Gamma_{2n+1}+1}2\right)\ .
\label{eq:massmatrix}
\end{equation}
In order for this matrix $M_r$ not to vanish, 
$\Gamma_{a_1a_2\cdots a_r}$ must be $\gamma_5$-diagonal and hence $r$ must be 
even\footnote{
Since $\Gamma_a (a=1,2,\cdots,2n)$ flip the chirality $\Gamma_{2n+1}$,
the rank-$r$ anti-symmetric tensor $\Gamma$-matrices $\Gamma_{a_1a_2\cdots a_r}$
are $\gamma_5$-diagonal for even $r$ and $\gamma_5$-off-diagonal for odd $r$.}.
Moreover, in order for the mass term $\psi^{\rm T}(C\,M_r)\psi$ not to vanish, 
the matrix $C M_r$ should also be $\gamma_5$-diagonal, 
and hence $C$ as well as $M_r$ must be $\gamma_5$-diagonal. 
From Eq.~(\ref{eq:C-matrix}), we find that $C$ is $\gamma_5$-diagonal
only for even $n$; that is, the cases where 
$SO(N=2n)$ spinor representations are either
real for $n\equiv0\ (\mbox{mod.4})$, 
or pseudo-real for $n\equiv2\ (\mbox{mod.4})$. 
For those cases, the mass matrix 
$M_r$ in (\ref{eq:massmatrix}) is still traceless since both 
$\tr\left[\Gamma_{a_1a_2\cdots a_r}\right]=0$ and 
$\tr\left[\Gamma_{a_1a_2\cdots a_r}\Gamma_{2n+1}\right]=0$ 
hold, and can give non-trivial constraint on the fermion mass 
that the sum of mass eigenvalues should vanish. 
Note, therefore, that the traceless condition on the 
mass eigenvalues is valid except only for $N=4n+2$ cases where 
the $SO(N)$ spinor representation is complex.

\section{Addition-type subgroups}
\label{Sec:Additon-type-subgroups}

We can write down 
the form of the auxiliary scalar VEV which breaks $SO(N)$ into 
an addition-type subgroup $H$:
\begin{align}
\langle\Phi^\dagger_{\bfR_k}\rangle= C\cdot V_{SO(N)\to H},
\end{align}
where we have factored out 
the charge conjugation matrix $C$ from the VEV
$\langle\Phi_{\bfR_k}^\dag\rangle$ in conformity with
Eq.~(\ref{eq:Yukawa}), 
so $V_{SO(N)\to H}$ corresponds to the fermion mass matrix $M_r$ 
to which the ``traceless condition'' can be applied.
$SO(N)$ spinor representations has $N$ modulo 8 structure.
$H$ represents one of the little group of $SO(N)$.
From Tables~\ref{tab:SO2n-maximal-subgroups-addition-type}
and \ref{tab:SO2n+1-maximal-subgroups-addition-type}
and by using the
expression of the gamma matrices in Eq.~(\ref{Eq:Gamma-matrix}),
we find the following VEVs in general; \ 
for $\VEV{\Phi^\dagger_{\bfR_0}}$, 
\begin{align}
\begin{array}{rclcl}
 V_{SO(4\ell)\to SO(2\ell)\times SO(2\ell)}&
 \propto&\Gamma_{1\cdots2\ell}
 & \propto&\sigma_0^{\otimes \ell-1}\otimes \sigma_3\otimes \sigma_0^{\otimes \ell},\\
 V_{SO(4\ell+1)\to SO(2\ell)\times SO(2\ell+1)}&
 \propto&\Gamma_{1\cdots2\ell}
 & \propto&\sigma_0^{\otimes \ell-1}\otimes \sigma_3\otimes \sigma_0^{\otimes \ell},\\
 V_{SO(4\ell+2)\to SO(2\ell+1)\times SO(2\ell+1)}&
 \propto&\Gamma_{1\cdots2\ell+1}
 & \propto&\sigma_0^{\otimes \ell}\otimes \sigma_1^{\otimes \ell+1},\\
 V_{SO(4\ell+3)\to SO(2\ell+1)\times SO(2\ell+2)}&
 \propto&\Gamma_{1\cdots2\ell+1}
 & \propto&\sigma_0^{\otimes \ell}\otimes \sigma_1^{\otimes \ell+1},\\
\end{array}
\end{align}
for $\VEV{\Phi^\dagger_{\bfR_q'}}$ $(q=1,2,\cdots, \left[\frac{N+1}{8}\right])$,
\begin{align}
\begin{array}{rclcl}
 V_{SO(4\ell+1)\to SO(2\ell+1-4q)\times SO(2\ell+4q)}&
 \propto&\Gamma_{1\cdots2\ell+1-4q}
 & \propto&\sigma_0^{\otimes \ell-2q}\otimes \sigma_1^{\otimes \ell+2q},\\
 V_{SO(4\ell+3)\to SO(2\ell+2-4q)\times SO(2\ell+1+4q)}&
 \propto& \Gamma_{1\cdots2\ell+2-4q} 
 & \propto&\sigma_0^{\otimes \ell-2q}\otimes \sigma_3\otimes \sigma_0^{\otimes \ell+2q},\\
\end{array} 
\end{align}
for $\VEV{\Phi^\dagger_{\bfR_p}}$ $(p=1,2,\cdots, \left[\frac{N}{8}\right])$,
\begin{align}
\begin{array}{rclcl} 
 V_{SO(4\ell)\to SO(2\ell-4p)\times SO(2\ell+4p)}&
 \propto&\Gamma_{1\cdots2\ell-4p}
 & \propto&\sigma_0^{\otimes \ell-1-2p}\otimes \sigma_3\otimes \sigma_0^{\otimes \ell+2p},\\
 V_{SO(4\ell+1)\to SO(2\ell-4p)\times SO(2\ell+1+4p)}&
 \propto&\Gamma_{1\cdots2\ell-4p}
 & \propto&\sigma_0^{\otimes \ell-1-2p}\otimes \sigma_3\otimes \sigma_0^{\otimes \ell+2p},\\
 V_{SO(4\ell+2)\to SO(2\ell+1-4p)\times SO(2\ell+1+4p)}&
 \propto&\Gamma_{1\cdots2\ell+1-4p}
 & \propto&\sigma_0^{\otimes \ell-2p}\otimes \sigma_1^{\otimes \ell+1+2p},\\
 V_{SO(4\ell+3)\to SO(2\ell+1-4p)\times SO(2\ell+2+4p)}&
 \propto&\Gamma_{1\cdots2\ell+1-4p}
 & \propto&\sigma_0^{\otimes \ell-2p}\otimes \sigma_1^{\otimes \ell+1+2p}.\\
\end{array} 
\end{align}
It should be understood that the chiral projection matrix 
\begin{equation}
\calP := \frac{\Gamma_{2n+1}\pm1}2
\end{equation}
is multiplied to these expressions for $V_{SO(N)\to H}$ when $N$ is even 
$N=2n$.
Any VEV of these is proportional to 
$\Gamma_{1\cdots r}$ of a certain rank $r$, 
so the quadratic mass matrix of the $SO(N)$ spinor fermion is 
found to be 
\begin{align}
 \langle\Phi_{\bfR_k}\rangle\langle\Phi_{\bfR_k}^\dag\rangle 
 &\propto\calP \Gamma^\dagger_{1\cdots r}C^\dagger C\Gamma_{1\cdots r} \calP
 \propto{\bf1}_{2^{\left[\frac{N}{2}\right]}} \calP
\end{align}
with understanding that $\calP=1$ for $SO(N)$ with odd $N=2n+1$ case.
That is, all the $d=2^{\left[\frac{N-1}{2}\right]}$ 
components of $\psi$ obtain a degenerate 
mass square $m^2$ realizing the minimum of $F(x, M_{\bfR_k}^2)$ for all
cases. Therefore the global minimum of the potential is realized by the
VEV of $\Phi_{\bfR_k}$ for which the coupling constant 
$G_{\bfR_k}$ is the strongest.
So the symmetry breaking pattern is determined depending on which
coupling constant $G_{\bfR_p}$ is the strongest: We find the following
breaking pattern.
For even $N=2n$,
\begin{align}
SO(N=2n)& \quad \rightarrow\quad 
\begin{cases} 
SO(n)\times SO(n) & \hbox{for}\
 G_{\bfR_0} \hbox{:\ strongest}\\
SO(n+4)\times SO(n-4) & \hbox{for}\
 G_{\bfR_1} \hbox{:\ strongest}\\
SO(n+8)\times SO(n-8) & \hbox{for}\
 G_{\bfR_2} \hbox{:\ strongest}\\
SO(n+12)\times SO(n-12) & \hbox{for}\
 G_{\bfR_3} \hbox{:\ strongest}\\
\hspace{10em}\vdots &\\
\end{cases}.
\end{align}
For odd $N=2n+1$,
\begin{align}
SO(N=2n+1)& \quad \rightarrow\quad 
\begin{cases} 
SO(n+1)\times SO(n) & \hbox{for}\
 G_{\bfR_0} \hbox{:\ strongest}\\
SO(n+4)\times SO(n-3) & \hbox{for}\
 G_{\bfR_1'} \hbox{:\ strongest}\\
SO(n+5)\times SO(n-4) & \hbox{for}\
 G_{\bfR_1} \hbox{:\ strongest}\\
SO(n+8)\times SO(n-7) & \hbox{for}\
 G_{\bfR_2'} \hbox{:\ strongest}\\
SO(n+9)\times SO(n-8) & \hbox{for}\
 G_{\bfR_2} \hbox{:\ strongest}\\
\hspace{10em}\vdots &\\
\end{cases}.
\end{align}

\section{Product-type subgroups}
\label{Sec:Product-type-subgroups}

Next, we discuss ``product-type'' little group cases. 
From Table~\ref{tab:SO-maximal-subgroups-product-type}, 
we see that this ``product-type'' subgroups are 
further classified into three sub-types;  i.e., 
``orthogonal-type,'' ``symplectic-type,'' and
``unitary-type,'' subgroups. 
More explicitly, here we consider 
$SO(mn)\to SO(m)\times SO(n)$(S)
$(m,n\in\mathbb{Z}_{\geq3\backslash4})$,
and $SO(4mn)\to \USp(2m)\times \USp(2n)$(S)
$(m,n\in\mathbb{Z}_{\geq1})$, and 
$SO(2n)\to SU(n)\times U(1)$(R)
$(n\in\mathbb{Z}_{\geq2})$.
Note that $\USp(2)\simeq SU(2)$
and $SO(4)\simeq SU(2)\times SU(2)$ lead to
$SO(4n)\supset\USp(2)\times\USp(2n)
\simeq SU(2)\times\USp(2n)
\supset SU(2)\times SU(2)\times SO(n)
\simeq SO(4)\times SO(n)$.
Thus, $SO(4)\times SO(n)$ is not a maximal subgroup of $SO(4n)$.

For the branching rules for the cases
$SO(mn)\supset SO(m)\times SO(n)$
$(m,n\in\mathbb{Z}_{\geq3\backslash4})$
and $SO(4mn)\supset \USp(2m)\times \USp(2n)$
$(m,n\in\mathbb{Z}_{\geq1})$,
many examples are listed in the tables in Ref.~\cite{Yamatsu:2015gut}.

As we will see below, we find that 
only $SO(9)\to SO(3)\times SO(3)$(S) and 
$SO(16)\to \USp(4)\times \USp(4)$(S),
$SO(4)\to SU(2)\times U(1)$(R)
breaking cases can satisfy the
global minimum condition in appropriate NJL coupling constant region
because all the fermion masses are degenerate for those breaking.

\subsection{Orthogonal-type}
\label{Sec:Product-Orthogonal}

For the breaking $SO(mn)\supset SO(m)_A\times SO(n)_B$,  
the $SO(mn)$ vector index 
$a$ ($1\leq a\leq mn$) is identified with a pair $\{iI\}$ of 
$SO(m)_A$ vector index $i$ ($1\leq i\leq m$) and $SO(n)_B$ vector index
$I$ ($1\leq I\leq n$); more explicitly, the $SO(mn)$ Gamma matrices
$\Gamma_a$ are denoted in both way as 
\begin{align}
\Gamma_{\{iI\}}\ \leftrightarrow \ \Gamma_{a} \ (\hbox{with}\  a=n(i-1)+I)\,.
\end{align}
Then the $SO(m)_A$ and $SO(n)_B$ generators $J_{ij}^A$ and $J_{IJ}^B$ 
are given by the following linear combinations of the 
$SO(mn)$ generators
$\propto\Gamma_{ab}=\Gamma_a\Gamma_b=-\Gamma_b\Gamma_a$ (with $a\not=b$):
\begin{align}
 J_{ij}^A=\frac{1}{2\sqrt{2}i}
 \sum_{I,J=1}^{n}\delta^{IJ}\Gamma_{\{iI\}\{jJ\}},\ \ \
 J_{IJ}^B=\frac{1}{2\sqrt{2}i}
 \sum_{i,j=1}^{m}\delta^{ij}\Gamma_{\{iI\}\{jJ\}}.
\label{Eq:SOmn-SOm-SOn-generators}
\end{align}
The quadratic Casimir operator of $SO(m)_A$ is calculated as
\begin{align}
 \sum_{j>i=1}^m(J_{ij}^A)^2
 &=-\frac1{(2\sqrt{2}i)^2}
 \sum_{j>i=1}^m\sum_{I,J=1}^{n}\Gamma_{\{iI\}\{jI\}}\Gamma_{\{iJ\}\{jJ\}} 
 =-\frac18
 \left( -{\bf1}\times n\times\frac{m(m-1)}2 + 2M_4 \right), \nn
 M_4 &:=\sum_{j>i=1}^m \sum_{J>I=1}^{n}\Gamma_{\{iI\}\{jI\}}\Gamma_{\{iJ\}\{jJ\}},\ \ \
\label{Eq:CasimirOperator}
\end{align}
where the first terms $\propto\bf1$ come from $I=J$ terms and the second
term $2M_4$ from $I\not=J$, which is a (linear combination of) rank-4 
anti-symmetric tensor 
$\Gamma_{\{iI\}\{jI\}\{iJ\}\{jJ\}}$ since $i\not=j, I\not=J$. 
Note that since $SO(m)_A$ generators $J_{ij}^A$ are $SO(n)_B$-invariant 
by construction and so the Casimir operator $(J_{ij}^A)^2$ is both
invariant under $SO(m)_A$ and $SO(n)_B$. But the first term
$\propto\bf1$ is $G=SO(mn)$-invariant, so that 
the rest rank-4 tensor part $M_4$
gives an $H=SO(m)_A\times SO(n)_B$ singlet component in the rank-4
$SO(mn)$ tensor $\Gamma_{abcd}$. Note also that the quadratic Casimir of
the other factor group $SO(n)_B$ reproduce the same rank-4 $H$-singlet
$M_4$ with minus sign as 
\begin{align}
 \sum_{j>i=1}^m(J_{IJ}^B)^2
 &=-\frac1{(2\sqrt{2}i)^2}
 \sum_{J>I=1}^n\sum_{i,j=1}^{m}\Gamma_{\{iI\}\{iJ\}}\Gamma_{\{jI\}\{jJ\}} 
 =-\frac18
 \left( -{\bf1}\times m\times\frac{n(n-1)}2 - 2M_4 \right). 
\label{Eq:CasimirOperatorB}
\end{align}
This should be so because the rank-4 $H$-singlet is unique in the 
$SO(mn)$ rank-4 tensor $\Gamma_{abcd}$, or more technically, because 
\begin{eqnarray}
\Gamma_{\{iI\}\{jI\}}\Gamma_{\{iJ\}\{jJ\}}=
\Gamma_{\{iI\}}\Gamma_{\{jI\}}\Gamma_{\{iJ\}}\Gamma_{\{jJ\}}=
-\Gamma_{\{iI\}}\Gamma_{\{iJ\}}\Gamma_{\{jI\}}\Gamma_{\{jJ\}}=
-\Gamma_{\{iI\}\{iJ\}}\Gamma_{\{jI\}\{jJ\}}. \nonumber
\end{eqnarray}

First, we show that the $SO(9)\to SO(3)_A\times SO(3)_B$ breaking VEV of
the $SO(9)$ {\bf 126}\ 
(rank-4 AS tensor) scalar field $\Phi_{\bfR_0}$ generates
a common mass for
the $SO(9)$ spinor fermion ${\bf 16}$. 
The $SO(9)$ spinor ${\bf 16}$ is decomposed into two representations
under $SO(9)\supset SO(3)_A\times SO(3)_B$ as 
\begin{align}
{\bf 16}=({\bf 4,2})\oplus({\bf 2,4}).
\end{align}
(See e.g., Ref.~\cite{Yamatsu:2015gut}.)
With $SO(3)$ generator denoted in 3-vector notation 
$\vec{J}=(J_{23}, J_{31}, J_{12})$ as usual, the quadratic Casimir
operator $(\vec{J}^A)^2$ of $SO(3)_A$ subgroup possesses the following 
eigenvalues on the $SO(9)$ spinor ${\bf 16}$:
\begin{align}
(\vec{J}^A)^2
 \left(
 \begin{array}{c}
  |({\bf 4,2})\rangle\\
  |({\bf 2,4})\rangle\\
 \end{array}
 \right)
 =
 \left(
 \begin{array}{c}
  C_2({\bf4})|({\bf 4,2})\rangle\\
  C_2({\bf2})|({\bf 2,4})\rangle\\
 \end{array}
 \right)
 =
 \left(
 \begin{array}{c}
  \frac{15}8\, |({\bf 4,2})\rangle\\[.8ex]
  \frac38\, |({\bf 2,4})\rangle\\
 \end{array}
 \right),
\end{align}
where the $SO(3)$ Casimir eigenvalues
$C_2({\bf 4})=15/8, C_2({\bf 2})=3/8$ 
here 
are halves of the $SU(2)$ ones $j(j+1)$ for spin-$j$ representation 
listed in Ref.~\cite{Yamatsu:2015gut}
because of our convention $T({\bf3})=1$ for $SO(3)$.
The quadratic Casimir operators (\ref{Eq:CasimirOperator}) and 
(\ref{Eq:CasimirOperatorB}) in this $m=n=3$ case 
are given by 
\begin{align}
 (\vec{J}^A)^2=
 -\frac{1}{8}
 \left(-{\bf 1}_{16}\times3\times3+2M_4\right),\ \ \
 (\vec{J}^B)^2=
 -\frac{1}{8}
 \left(-{\bf 1}_{16}\times3\times3-2M_4\right),
\end{align}
with the rank-4 tensor part $M_4$ which more explicitly reads by 
using $SO(9)$ vector index 
$a = 2i+I$ in place of $\{iI\}$ notation 
$(a=1,2,\cdots,9; \ i,I=1,2,3)$,
\begin{align}
-M_4:=\Gamma_{1245}+\Gamma_{1346}+\Gamma_{2356}+\Gamma_{1278}+\Gamma_{1379}
 +\Gamma_{2389}+\Gamma_{4578}+\Gamma_{4679}+\Gamma_{5689}.
\label{eq:SO33singlet}
\end{align}
Since the $M_4$ is an $SO(3)_A\times SO(3)_B$-singlet in the rank-4
tensor $\Gamma_{abcd}$ as noted above, 
the  fermion mass matrix is proportional to 
$M_4$ and acts on the $SO(9)$ spinor ${\bf16}$ as
\begin{align}
-v M_4\left(
 \begin{array}{c}
  |({\bf 4,2})\rangle\\
  |({\bf 2,4})\rangle\\
 \end{array}
 \right) 
&=
 \frac12 v \left(8(\vec{J}^A)^2- 9\times{\bf 1}_{16}\right)
 \left(
 \begin{array}{c}
  |({\bf 4,2})\rangle\\
  |({\bf 2,4})\rangle\\
 \end{array}
 \right)
 =
 \left(
 \begin{array}{rl}
  +3v&\hspace{-.7em} |({\bf 4,2})\rangle\\
  -3v&\hspace{-.7em} |({\bf 2,4})\rangle\\
 \end{array}
 \right).
\label{eq:SO9to33} 
\end{align}
This leads to the fermion mass square matrix proportional to 
$(M_4^\dag M_4) \propto{\bf 1}_{16}$.
Thus, we find that all the fermion states in 
$({\bf 4,2})$ and $({\bf 2,4})$ of $SO(3)_A\times SO(3)_B\subset SO(9)$
get the same mass from 
the condensation of $SO(3)_A\times SO(3)_B$-singlet 
combination (\ref{eq:SO33singlet}) of rank-4 $\Gamma_{abcd}$. 
Note, however, that we could have obtained this result (\ref{eq:SO9to33}) 
more quickly from the ``traceless condition'' of the fermion mass 
matrix $M_4$ since it can be applied to $SO(9)$ case with odd $N=9$ 
and the 
$H$-irreducible sectors $({\bf 4,2})$ and $({\bf 2,4})$ have the 
same dimensions 8.

One may wonder whether for other cases such as
$SO(15)\to SO(3)_A\times SO(5)_B$ the VEV of the scalar field 
can also realize the global minimum or not.
For example, for $SO(15)\to SO(3)_A\times SO(5)_B$, the branching rules
of the $SO(15)$ spinor representation is given by
\begin{align}
{\bf 128}={\bf (6,4)}\oplus{\bf (4,16)}\oplus{\bf (2,20)}.
\end{align}
The eigenvalues of $SO(3)_A$ and $SO(5)_B$ Casimir operators 
$(\vec{J}^A)^2$ and $(\vec{J}^B)^2$   
of the $SO(15)$ spinor ${\bf 128}$ are
given by 
\begin{align}
(\vec{J}^A)^2
 \left(
 \begin{array}{l}
  |({\bf 6,4})\rangle\\
  |({\bf 4,16})\rangle\\
  |({\bf 2,20})\rangle\\
 \end{array}
 \right)
 =
 \frac18
 \left(
 \begin{array}{l}
 35\, |({\bf 6,4})\rangle\\
 15\, |({\bf 4,16})\rangle\\
 \ 3\, |({\bf 2,20})\rangle\\
 \end{array}
 \right), \\
(\vec{J}^B)^2
 \left(
 \begin{array}{l}
  |({\bf 6,4})\rangle\\
  |({\bf 4,16})\rangle\\
  |({\bf 2,20})\rangle\\
 \end{array}
 \right)
 =
 \frac18
 \left(
 \begin{array}{l}
 10\, |({\bf 6,4})\rangle\\
 30\, |({\bf 4,16})\rangle\\
 42\, |({\bf 2,20})\rangle\\
 \end{array}
 \right), 
\end{align}
where use has been made of,
for $SO(3)$, $C_2({\bf 6})=35/8, C_2({\bf 4})=15/8, C_2({\bf 2})=3/8$;
for $SO(5)$, $C_2({\bf 20})=21/4, C_2({\bf 16})=15/4, C_2({\bf 4})=5/4$.
(See e.g., Ref.~\cite{Yamatsu:2015gut}.)
From Eq.~(\ref{Eq:CasimirOperator}) with $m=3, n=5$, 
we find the quadratic Casimir operators for the subgroups 
$SO(3)_A$ and $SO(5)_B$:
\begin{align}
 (\vec{J}^A)^2=
 -\frac{1}{8}
 \left(-{\bf 1}_{128}\times5\times3+2M_4\right),\ \ \
 (\vec{J}^B)^2=
 -\frac{1}{8}
 \left(-{\bf 1}_{128}\times3\times10-2M_4\right),
\end{align}
with 
$ M_4=\sum_{j>i=1}^{3}\sum_{J>I=1}^{5}
 \Gamma_{\{iI\}\{jI\}\{iJ\}\{jJ\}}$. 
The fermion mass is proportional to the $SO(3)_A\times SO(5)_B$-singlet $M_4$
which acts on the $SO(15)$-spinor ${\bf 128}$ as 
either $-(8(\vec{J}^A)^2-15\,{\bf1}_{128})/2$ or 
$(8(\vec{J}^B)^2-30\,{\bf1}_{128})/2$. In either way, we find
\begin{align}
-vM_4\left(
 \begin{array}{l}
  |({\bf 6,4})\rangle\\
  |({\bf 4,16})\rangle\\
  |({\bf 2,20})\rangle\\
 \end{array}
 \right)
 &\propto 
 \left(
 \begin{array}{rl}
  10v&\hspace{-.7em}|({\bf 6,4})\rangle\\
  \ \ 0&\hspace{-.7em}|({\bf 4,16})\rangle\\
  -6v&\hspace{-.7em}|({\bf 2,20})\rangle\\
 \end{array}
 \right).
\end{align}
This leads to $(M_4^\dag M_4)\not\propto{\bf 1}$.
Thus, we find that the fermion states in
$|({\bf 6,4})\rangle$, $|({\bf 4,16})\rangle$, and $({\bf 2,20})\rangle$
of $SO(3)_A\times SO(5)_B\subset SO(15)$
obtain different masses from the rank-4 tensor $\Gamma_{abcd}$
condensation.
Note here also 
that the fermion mass matrix $-vM_4$ satisfies the traceless property,  
$\tr M_4=0$; indeed
\begin{equation}
\tr(-vM_4) = 10v\times(6\times4) + 0\times(4\times16)
 -6v\times(2\times20)=0.
\end{equation}
This again confirms the validity of our traceless condition of the
fermion mass matrix. In this case, however, the original spinor is
decomposed into three $H$-irreducible components so that the traceless
condition alone cannot determine the three unknown mass eigenvalues even
aside from the overall scale.

In general, when the breaking $SO(mn) \to SO(m)\times SO(n) (m,n\geq3)$
occurs, the $SO(mn)$ spinor representation is decomposed into at least
two different $H$-irreducible representations. 
(Several examples can be found in Ref.~\cite{Yamatsu:2015gut}.)
The $SO(9)\to SO(3)\times SO(3)$ case is rather exceptional where 
only two $H$-irreducible components appear and they have the 
same dimensions. Then, by the traceless condition, the mass matrix 
$M_4$ turns to have two mass eigenvalues $\pm3v$ with same magnitude 
and opposite sign, so that 
the fermion quadratic mass matrix $(M_4^\dag M_4)$ was proportional to 
an identity.
For general $m\not=n$ cases, however, the dimensions of the
$H$-irreducible components 
are mutually different and so, even for the case where only two
$H$-irreducible components appear, 
the traceless arguments for the mass matrix $M_r$  
gives two different eigenvalues for those $H$-irreducible components, so 
$(M_4^\dag M_4)\not\propto{\bf1}$.
Even for $m=n$ cases, the $SO(mn)$ spinor splits into 
more than or equal to four $H$-irreducible representations for 
$m=n\geq5$, 
and one easily see by the argument of quadratic Casimir and tracelessness 
that the mass matrix $M_4$ has eigenvalues of at least two different 
absolute values, so that $(M_4^\dag M_4)\not\propto{\bf1}$.
Therefore, the $SO(mn)\to SO(m)\times SO(n) (m,n\geq3)$ breaking
VEV cannot be the global minimum of the potential in the $SO(mn)$
NJL-type model except for the $m=n=3$ case.

\subsection{Symplectic-type}
\label{Sec:Product-Symplectic}

Here we discuss the $SO(4mn)\to\USp(2m)\times\USp(2n)$ breaking
case. 
A close parallelism holds between 
the previous breaking $SO(mn)\to SO(m)\times SO(n)$ in 
Sec.~\ref{Sec:Product-Orthogonal} and the present one 
$SO(4mn)\to\USp(2m)\times\USp(2n)$. 

From the tables in Ref.~\cite{Yamatsu:2015gut} and additional searches,
we find that under the subgroup $H=\USp(2m)\times\USp(2n)\subset SO(4mn)$ 
except the $m=1,n=2$ and $m=n=2$ cases, the $SO(4mn)$ spinor
representation is decomposed into two essentially different
$H$-irreducible representations. So the quadratic mass matrix
$M_r M^\dagger_r$ of the fermion cannot be 
proportional to identity. Thus, they cannot realize the global minimum of
the effective potential.

Consider the first exceptional case of $m=1,n=2$. 
For $\USp(2)\times\USp(4)\subset SO(8)$
the branching rules of the $SO(8)$ spinor representations $\bfr={\bf 8_s}$
and $\bfr'={\bf8_c}$ (both are real)
are given by
\begin{align}
 {\bf 8_s}=({\bf 2,4}), \qquad 
 {\bf 8_c}=({\bf 2,4});
\end{align}
that is, they are also $H$-irreducible. In this case, however, 
the symmetric tensor product of the $SO(8)$ spinor gives 
$({\bf8}\times{\bf8})_S={\bf35}(\bfR_0)+{\bf1}(\bfR_1)$ (for both cases 
${\bf 8_s}$ and ${\bf 8_c}$), but $\Phi_{\bfR_0}$ does not have a 
singlet of $H=\USp(2)\times\USp(4)\subset SO(8)$ and $\Phi_{\bfR_1}$ is $SO(8)$-singlet. 
So the breaking $SO(8)\to \USp(2)\times\USp(4)$ actually does not occur 
in this case.

Next consider the second exceptional case $m=n=2$.  This case, 
the two $SO(16)$ irreducible spinor representations 
$\bfr={\bf128}$ and $\bfr'={\bf128'}$ 
have different branching rules under $\USp(4)\times\USp(4)\subset 
SO(16)$ as
\begin{align}
{\bf 128}={\bf (14,1)}\oplus{\bf (10,5)}\oplus{\bf (5,10)}\oplus{\bf
 (1,14)},\ \ \
{\bf 128'}={\bf (16,4)}\oplus{\bf (4,16)}.
\end{align}
The pair condensation of the ${\bf 128}$ spinor fermion cannot realize
the global minimum of the potential since the fermion mass eigenvalues 
are not degenerate in this case. This can be shown as follows. 

The traceless condition holds in this case $SO(16)$ possessing real
spinor. 
For the case of {\bf 128} spinor which decomposes into more than two 
$H$-irreducible components, the traceless condition alone is not enough 
to calculate the fermion mass eigenvalues.
Since the $\USp(4)\times\USp(4)$-singlet 
exists in the channel $\Phi_{\bfR_1}$ (rank-4 tensor), the eigenvalue of
the mass matrix $M_4$ in the rank-4 tensor can be calculated by
computing the $\USp(4)$ quadratic Casimir $C_2$ which must be a linear
combination of $G$-singlet matrix ${\bf 1}\in\Phi_{\bfR_2}$ and 
the $H$-singlet $M_4\in\Phi_{\bfR_1}$. Hence we have 
$M_4\propto C_2 - x{\bf1}$:  
\begin{align}
M_4\left(
 \begin{array}{l}
  |({\bf 14,1})\rangle\\
  |({\bf 10,5})\rangle\\
  |({\bf 5,10})\rangle\\
  |({\bf 1,14})\rangle\\
 \end{array}
 \right)
 &\propto 
 \left(
 \begin{array}{l}
  \left(5-x\right)|({\bf 14,1})\rangle\\
  \left(3-x\right)|({\bf 10,5})\rangle\\
  \left(2-x\right)|({\bf 5,10})\rangle\\
  \left(0-x\right)|({\bf 1,14})\rangle\\
 \end{array}
 \right)
 =
 \left(
 \begin{array}{l}
  +\frac{5}{2}|({\bf 14,1})\rangle\\
  +\frac{1}{2}|({\bf 10,5})\rangle\\
  -\frac{1}{2}|({\bf 5,10})\rangle\\
  -\frac{5}{2}|({\bf 1,14})\rangle\\
 \end{array}
 \right),
\end{align}
where the $\USp(4)$ Casimir eigenvalues $C_2({\bf 14})=5$,
$C_2({\bf 10})=3$, $C_2({\bf 5})=2$, $C_2({\bf 1})=0$ have been used and, 
in going to the last expression,  
the $G$-singlet coefficient $x$ was found to be $x=5/2$ by using the 
traceless condition:
\begin{equation}
5\times14+3\times(10\times5)+2\times(5\times10)+0\times14 - 
128x=0 \quad \rightarrow\quad  x=\frac{320}{128}=\frac52.
\end{equation}
Note that we were able to skip the laborious explicit identification 
of the $\USp(4)$ generators in terms of the original $SO(16)$ generators 
$\Gamma_{ab}$.  

For the case of the spinor $\bfr'={\bf 128'}$,  on the other hand, 
the traceless condition is enough to determine the (ratio of) the mass 
eigenvalues since $\bfr'={\bf 128'}$ has only 
two $H$-irreducible components under $H=\USp(4)\times\USp(4)\subset SO(16)$.
Again the $H$-singlet is in the $\Phi_{\bfR_1}$ (rank-4 tensor) so 
the mass matrix $M_4$ must be traceless: 
\begin{align}
M_4 \left(
 \begin{array}{l}
  |({\bf 16,4})\rangle\\
  |({\bf 4,16})\rangle\\
 \end{array}
 \right)
 \propto 
 \left(
 \begin{array}{l}
  +1\,|({\bf 16,4})\rangle\\
  -1\,|({\bf 4,16})\rangle\\
 \end{array}
 \right).
\end{align}
Thus, their quadratic mass matrix is proportional to
identity and can realize the global minimum.

For $\USp(2)\times\USp(8)\subset SO(16)$, 
the branching rules of the $SO(16)$ spinor representations are given by
\begin{align}
{\bf 128}={\bf (5,1)}\oplus{\bf (3,27)}\oplus{\bf (1,42)},\ \ \
{\bf 128'}={\bf (4,8)}\oplus{\bf (2,48)}.
\end{align}
Again the $\USp(2)\times\USp(8)$-singlet is in the 
$\Phi_{\bfR_1}$ (rank-4 tensor) 
both for ${\bf 128}$ and ${\bf 128'}$ spinors. The traceless condition and
quadratic Casimir lead to
\begin{align}
M_4\left(
 \begin{array}{l}
  |({\bf 5,1})\rangle\\
  |({\bf 3,27})\rangle\\
  |({\bf 1,42})\rangle\\
 \end{array}
 \right)
 &\propto 
 \left(
 \begin{array}{l}
  \left(6-\frac{3}{2}\right)|({\bf 5,1})\rangle\\
  \left(2-\frac{3}{2}\right)|({\bf 3,27})\rangle\\
  \left(0-\frac{3}{2}\right)|({\bf 1,42})\rangle\\
 \end{array}
 \right)
 =
 \left(
 \begin{array}{l}
  +\frac{9}{2}|({\bf 5,1})\rangle\\
  +\frac{1}{2}|({\bf 3,27})\rangle\\
  -\frac{3}{2}|({\bf 1,42})\rangle\\
 \end{array}
 \right),\\
M_4\left(
 \begin{array}{l}
  |({\bf 4,8})\rangle\\
  |({\bf 2,48})\rangle\\
 \end{array}
 \right)
 &\propto 
 \left(
 \begin{array}{l}
  \left(\frac{15}{4}-\frac{3}{2}\right)|({\bf 4,8})\rangle\\
  \left(\frac{3}{4}-\frac{3}{2}\right)|({\bf 2,48})\rangle\\
 \end{array}
 \right)
 =
 \left(
 \begin{array}{l}
  +\frac{9}{4}|({\bf 4,8})\rangle\\
  -\frac{3}{4}|({\bf 2,48})\rangle\\
 \end{array}
 \right),
\end{align}
where the $\USp(2)\simeq SU(2)$ Casimir eigenvalues
$C_2({\bf 5})=6$, $C_2({\bf 3})=2$, $C_2({\bf 1})=0$,
$C_2({\bf 4})=15/4$, $C_2({\bf 2})=3/4$ have been used.
The pair condensation of the ${\bf 128}$ and ${\bf 128'}$ spinor fermion, 
therefore, cannot realize the global minimum of the potential.

\subsection{Unitary-type}
\label{Sec:Product-Unitary}

Here we consider the $SO(2n)\to SU(n)\times U(1)$ or $SU(n)$ $(n\geq2)$
breaking cases for even and odd $n$, respectively.
For $SO(4k)\to SU(2k)\times U(1)$ (even $n=2k$),
the spinor representations are real or pseudo-real representations;
for $SO(4k+2)\to SU(2k+1)\times U(1)$ (odd $n=2k+1$),
the spinor representations are complex representations.
For $SO(4k)$, the fermion mass matrix must satisfy the traceless
condition.
From the following discussion, we find that for only $n=2$, i.e., 
$SO(4)\simeq SU(2)\times SU(2)\to SU(2)\times U(1)$,
the pair condensation of the spinor fermion realizes the global minimum
of the potential.

To calculate the $SU(n)$ decomposition of the $SO(2n)$ spinor fermion, 
we here follow the well-known trick to rewrite the 
$SO(2n)$ gamma matrices $\Gamma_{a}$ into 
the creation and annihilation operators of $SU(n)$ fermion, 
$a_k^\dag, a^k (k=1,2,\cdots, n)$\cite{Mohapatra:1979nn}:
\begin{align}
\left\{
 \begin{array}{l}
 a_k^\dag:=\frac{1}{2}\left(\Gamma_{2k-1}+i\Gamma_{2k}\right)\\[1ex]
 a^{k}:=\frac{1}{2}\left(\Gamma_{2k-1}-i\Gamma_{2k}\right)\\
 \end{array}
 \right.
\  \leftrightarrow\
\left\{
 \begin{array}{l}
 \Gamma_{2k-1}=a_k^\dag+a^k\\[1ex]
 \Gamma_{2k}=\frac{1}{i}(a_k^\dag-a^k)
 \end{array}
 \right.\,. 
\end{align}
Then the $SO(2n)$ (reducible) $2^n$-dimensional spinor is represented 
in the form
\begin{align}
\ket{\psi} &= \ket0 \psi_0 +\half a_i^\dagger\ket0 \psi^i
+ \half a_i^\dagger a_j^\dagger\ket0 \psi^{ij}+ \cdots \nn
 &\hspace{1em}
 + \varepsilon^{i_1i_2\cdots i_{n-1}j}\frac1{(n-1)!}a_{i_1}^\dagger a_{i_2}^\dagger 
\cdots a_{i_{n-1}}^\dagger\ket0 \tilde\psi_{j}
+ a_{1}^\dagger a_{2}^\dagger\cdots a_{n}^\dagger\ket0 \tilde\psi_0,
\label{eq:spinor}
\end{align}
giving a decomposition into irreducible representations of $SU(n)$. 
The $SO(2n)$ chirality matrix ($\gamma_5$-analogue) is represented by
\begin{align}
 \Gamma_{\chi}&:=i^{n}\Gamma_{1}\Gamma_{2}\cdots\Gamma_{2n}
=\prod_{k=1}^n\left(1-2a_k^\dag a^k\right) 
=\prod_{k=1}^n(-1)^{a_k^\dag a^k}=(-1)^{\hat{N}},
\end{align}
with the total fermion number operator $\hat{N}:=\sum_k a_k^\dag a^k$.
So $SO(2n)$ irreducible chiral spinors $\ket{\psi_\pm}$ with
$\Gamma_\chi=\pm1$ are given  
by half sums of states in the RHS of Eq.~(\ref{eq:spinor}) possessing 
even and odd number of fermion excitations, respectively. 

Since the gamma matrices $\Gamma_{2k-1}$ are symmetric and 
$\Gamma_{2k}$ are anti-symmetric in the spinor 
basis of Eq.~(\ref{eq:spinor}),%
\footnote{%
In this spinor basis, 
the creation and annihilation operators $a^\dagger$ 
and $a$ are represented by $2\times2$ Pauli matrices as 
$\sigma^+=(\sigma_1+i\sigma_2)/2$ and $\sigma^-=(\sigma_1-i\sigma_2)/2$, respectively, in each 
$k$ sector. So, for the  
present case with $n$ species of fermions, 
the gamma matrices are represented as
\begin{equation}
 \Gamma_{2k-1}= \sigma_3^{\otimes k-1}\otimes \sigma_1\otimes  \sigma_0^{\otimes n-k}, 
\qquad  \Gamma_{2k}= \sigma_3^{\otimes k-1}\otimes \sigma_2\otimes \sigma_0^{\otimes n-k}. 
\end{equation} 
This representation of the $\Gamma_a$ matrices are different from 
that in Eq.~(\ref{Eq:Gamma-matrix}).
}
the charge conjugation matrix $C$ satisfying 
\begin{equation}
C^{-1} \Gamma_a^{\rm T} C = -\Gamma_a
\end{equation}
can be chosen to be the operator
\begin{equation}
C = i^n \prod_{k=1}^n \Gamma_{2k} = \prod_{k=1}^n (a_k^\dagger-a^k).
\label{eq:C}
\end{equation}

First, begin with the $n=2$ case, i.e.,
$SO(4)\simeq SU(2)\times SU(2)\to SU(2)(\times U(1))$. 
The branching rules of the $SO(4)$ spinor representations are given by
\begin{align}
{\bf (2,1)}={\bf (2)}(0),\ \ \
{\bf (1,2)}={\bf (1)}(+1)\oplus{\bf (1)}(-1).
\label{eq:4.30}
\end{align}
The symmetric tensor product of $SO(4)$ spinor gives
$({\bf (2,1)}\otimes{\bf (2,1)})_S={\bf (3,1)}(\bfR_0)$ and 
$({\bf (1,2)}\otimes{\bf (1,2)})_S={\bf (1,3)}(\bfR_0)$, which are 
respectively decomposed under under $SU(2)\subset SO(4)$ as 
${\bf (1,3)}(\bfR_0)$ has a singlet so the the pair condensation 
of the ${\bf (1,2)}$ spinor fermion can cause this breaking.

Now in the $SU(n)$ fermion representation, the $SO(4)$ chiral spinors 
with $\Gamma_\chi=\pm1$ are given by
\begin{align}
\left\{
 \begin{array}{l}
 |\psi_+\rangle=|0\rangle\psi_0
 +a_1^\dag a_2^\dag|0\rangle\tilde\psi_0\\
 |\psi_-\rangle=
  a_i^\dag |0\rangle\psi^i\\
 \end{array}
\right., 
\end{align}
where $\ket{\psi_+}$ and $\ket{\psi_-}$ correspond to ${\bf (1,2)}$ and 
${\bf (2,1)}$ in Eq.~(\ref{eq:4.30}), respectively. 
The bispinor representation $\bfR_0$ is now a rank-2 tensor 
$\Gamma_{a_1a_2}$ and so can be described by
the operators $\{a_i^\dag a_j^\dag, a_i^\dag a^j, a^i a^j\}$.
The $SU(2)$-singlet in $\bfR_0$ is clearly given by a linear combination 
$\{a_1^\dag a_2^\dag, a^1a^2\}$:
$c_1a_1^\dag a_2^\dag+c_2 a^2a^1$ with arbitrary coefficients $c_j(j=1,2)$.
The $SU(2)$-singlet mass matrix $M_2$ should also satisfy 
$M_2^\dag M_2=M_2M_2^\dag$ since it is diagonalizable as we have shown before, 
which further leads to a condition $|c_1|^2=|c_2|^2$. We choose the
phase convention of $a^1$ and $a^2$ such that $c_1=c_2=:m$ is satisfied with 
real $m$ and take $M_2$ as a Hermitian operator:
\begin{align}
 M_2=m\left(a_1^\dag a_2^\dag+a^2 a^1\right).
\end{align}
Acting this on the positive chiral spinor $|\psi_+\rangle$,
\begin{align}
 M_2|\psi_+\rangle 
 &=m\left(a_1^\dag a_2^\dag+a^2 a^1\right)
 \left( |0\rangle\psi_0
 +a_1^\dag a_2^\dag|0\rangle\tilde\psi_0 \right) \nonumber\\
 &=m \left(a_1^\dag a_2^\dag|0\rangle\psi_0
 +|0\rangle\tilde\psi_0\right)
 =\left(|0\rangle\ \ a_1^\dag a_2^\dag|0\rangle\right)
 \left(
\begin{array}{cc}
 0&m\\
 m&0\\
\end{array}
 \right)
 \left(
 \begin{array}{c}
 \psi_0\\
 \tilde\psi_0\\
 \end{array}
 \right).
\end{align}
We can diagonalize the mass matrix $M_2$ as
\begin{align}
M_2\left(
 \begin{array}{l}
  |({\bf 1})(+1)\rangle\\
  |({\bf 1})(-1)\rangle\\
 \end{array}
 \right)
 &=
 \left(
 \begin{array}{l}
  +m\,|({\bf 1})(+1)\rangle\\
  -m\,|({\bf 1})(-1)\rangle\\
 \end{array}
 \right). 
\end{align}
The pair condensation of the ${\bf (1,2)}$ spinor fermion thus yields a 
degenerate mass square eigenvalue for the spinor ${\bf (1,2)}$ and hence
realizes the global minimum of the potential.

Next, for $n=3$, i.e.,
$SO(6)\simeq SU(4)\to SU(3)\times U(1)$, 
the branching rules of the irreducible spinor representations
$\bfr={\bf 4}$ and $\bar\bfr={\bf \overline{4}}$ are given by 
\begin{align}
{\bf 4}={\bf (3)}(1)\oplus{\bf (1)}(-3),\ \ \
{\bf \overline{4}}={\bf (\overline{3})}(-1)\oplus{\bf (1)}(3).
\end{align}
The symmetric tensor product of $SO(6)$ spinor gives
$({\bf 4}\otimes{\bf 4})_S={\bf 10}(\bfR_0)$
$(({\bf \overline{4}}\otimes{\bf \overline{4}})_S={\bf \overline{10}}(\bfR_0))$.
${\bf 10}(\bfR_0)$ has a singlet under $SU(3)(\times U(1))\subset SU(4)$: 
\begin{align}
{\bf 10}={\bf (\overline{6})}(2)\oplus{\bf (3)}(-2)\oplus{\bf (1)}(-6),\ \ \
{\bf \overline{10}}={\bf (6)}(-2)\oplus{\bf (\overline{3})}(2)\oplus{\bf (1)}(6).
\end{align}
The $SO(6)$ chiral spinors are given in the $SU(n)$ fermion representation as
\begin{align}
\left\{
 \begin{array}{l}
  |{\bf {4}}\rangle=|\psi_-\rangle=
  a_i^\dag |0\rangle\psi^{i}
  +a_1^\dag a_2^\dag a_3^\dag |0\rangle\tilde\psi_0\\
 |{\bf \bar{4}}\rangle=|\psi_+\rangle=
   |0\rangle\psi_0
 +\frac{1}{2}\varepsilon^{ijk}a_i^\dag a_j^\dag|0\rangle\tilde\psi_k\\
 \end{array}
\right.. 
\end{align}
The mass matrix $M_3({\bf 10})$ 
($M_3({\bf \overline{10}})$) in the rank-3 tensor bispinor 
${\bf 10}({\bfR_0})$ (${\bf \overline{10}}({\bfR_0})$)
corresponds to an $SU(3)$ singlet operator:
\begin{align}
 M_3({\bf 10})=m a_1^\dag a_2^\dag a_3^\dag,\ \ \
 \left(M_3({\bf \overline{10}})=m a^3 a^2 a^1\right).
\end{align}
For odd $n$, the mass matrix $M_n$ and the charge conjugation matrix $C$
are both $\Gamma_\chi$-off diagonal, while $CM_n$ is $\Gamma_\chi$-diagonal.
So we directly have to calculate the fermion bilinear mass term 
$\psi_\pm^{\rm T}CM_n\psi_\pm$. 
Noting $\psi_\pm^{\rm T}$ is represented by the state
\begin{align}
\left\{
 \begin{array}{l}
 \langle{\bf {4}}^*|=\langle\psi_-^*|
 =\psi^i\langle0| a^i
 +\tilde \psi_0 \langle0|a^3 a^2 a^1\\
 \langle{\bf \bar{4}}^*|=\langle\psi_+^*|=
   \psi_0\langle0|
 +\frac{1}{2}\psi^{ij}\langle0|a^ja^i \\
 \end{array}
\right.\,,
\end{align}
and using the expression (\ref{eq:C}) of the charge conjugation operator $C$ 
for $n=3$,  the mass terms 
$\psi_\pm^{\rm T}CM_3\psi_\pm$ 
can be calculated as follows:
\begin{align}
\psi_-^{\rm T}CM_3\psi_- 
&=\langle{\bf 4}^*|C\,M_3({\bf \overline{10}})|{\bf 4}\rangle 
 =
 \langle\psi_-^*|
 \left(a_{1}^\dag-a^{1}\right)\left(a_{2}^\dag-a^{2}\right)
 \left(a_{3}^\dag-a^{3}\right)
 \left(m a^3 a^2 a^1\right) 
 |\psi_-\rangle\nn
 &=
 \langle\psi_-^*|
 \left(a_{1}^\dag-a^{1}\right)\left(a_{2}^\dag-a^{2}\right)
 \left(a_{3}^\dag-a^{3}\right)
 m \ket{0} \tilde\psi_0 
 =
 m\langle\psi_-^*|
 a_{1}^\dag a_{2}^\dag a_{3}^\dag
  \ket{0} \tilde\psi_0 
=  m\tilde\psi_0 \tilde\psi_0, 
\nonumber\\
\psi_+^{\rm T}CM_3\psi_+ 
&=\langle{\bf \bar{4}}^*|C M_3({\bf 10})|{\bf \bar{4}}\rangle 
=
 \langle\psi_+^*|
 \left(a_{1}^\dag-a^{1}\right)\left(a_{2}^\dag-a^{2}\right)
 \left(a_{3}^\dag-a^{3}\right)
 \left(m a_1^\dag a_2^\dag a_3^\dag\right) 
 |\psi_+\rangle\nn
 &=
 \langle\psi_+^*|
 \left(a_{1}^\dag-a^{1}\right)\left(a_{2}^\dag-a^{2}\right)
 \left(a_{3}^\dag-a^{3}\right)
 m a_1^\dag a_2^\dag a_3^\dag \ket{0} \psi_0 \nn
&=
 m\langle\psi_+^*|(-a^1a^2a^3)\,a_1^\dag a_2^\dag a_3^\dag \ket{0} \psi_0
=
 m\langle\psi_+^*| 0\rangle\psi_0
 =m \psi_0\psi_0\, .
\end{align}
That is, only the singlet $\tilde\psi_0$ in ${\bf \overline{4}}$ 
($\psi_0$ in ${\bf 4}$) gets the mass $m$ while 
the rest triplet ${\bf3}\,(\psi^i)\in{\bf \overline{4}}$ 
(${\bf \bar3}\,(\tilde\psi_i)\in{\bf \overline{4}}$) remains massless.
Thus, the pair condensation of the ${\bf 4}$ (${\bf \overline{4}}$) 
spinor fermion cannot
realize the global minimum of the potential.

This argument actually holds for all cases of odd $n=2\ell+1$, for which
the chiral spinor is complex 
\begin{align}
\ket{\psi_+} &= \ket{0}\psi_0+ \half a_i^\dagger a_j^\dagger\ket{0}\psi^{ij}+\cdots 
+\frac1{(n-1)!}\varepsilon^{i_1i_2\cdots i_{n-1}j}
a_{i_1}^\dagger a_{i_2}^\dagger\cdots a_{i_{n-1}}^\dagger\ket{0}
 \tilde\psi_j\, ,
\nn
\ket{\psi_-} &= a_i^\dagger\ket{0}\psi^i+ \half a_i^\dagger a_j^\dagger a_k^\dagger\ket{0}\psi^{ijk}
+\cdots 
+a_1^\dagger a_2^\dagger\cdots a_n^\dagger\ket0 \tilde\psi_0 \, ,
\end{align}
and the symmetric product of $SO(2n)=SO(4\ell+2)$ spinor contains 
the $SU(n)$ singlet only in the $\bfR_0$ (rank $n$ tensor). The mass operator 
$M_n$ is thus given by 
\begin{equation}
M_n= m a_1^\dagger a_2^\dagger\cdots a_n^\dagger\quad \text{or}\quad ma^n a^{n-1}\cdots a^2a^1.
\label{eq:R0mass}
\end{equation}
Then the fermion mass term $\psi_\pm^{\rm T}CM_n\psi_\pm$ is calculated
in quite the same way as above:
\begin{align}
\psi_-^{\rm T}CM_n\psi_- 
&=
 \langle\psi_-^*| C  \left(m a^n \cdots a^2 a^1\right) 
 |\psi_-\rangle 
 =
 \langle\psi_-^*|
 \left(\prod_{k=1}^n(a_k^\dag-a^k)\right)
 m \ket{0} \tilde\psi_0 \nn
&=
 m\langle\psi_-^*|
 a_{1}^\dag a_{2}^\dag \cdots a_{n}^\dag
  \ket{0} \tilde\psi_0 
=  m\tilde\psi_0 \tilde\psi_0, 
\nn
\psi_+^{\rm T}CM_n\psi_+ 
&=
 \langle\psi_+^*| C
 \left(m a_1^\dag a_2^\dag \cdots a_n^\dag\right) 
 |\psi_+\rangle 
=
 \langle\psi_+^*|
 \left(\prod_{k=1}^n(a_k^\dag-a^k)\right)
 m a_1^\dag a_2^\dag \cdots a_n^\dag \ket{0} \psi_0 
\nn
&=
 m\langle\psi_+^*|(-1)^n a^1a^2\cdots a^n\,a_1^\dag a_2^\dag \cdots a_n^\dag \ket{0} \psi_0
=
 (-1)^{n+\ell}m\langle\psi_+^*| 0\rangle\psi_0
 =(-1)^{\ell+1}m \psi_0\psi_0\, .
\end{align}
In either case of $\pm$ chiral spinors, only the singlet component 
$\tilde\psi_0$ or $\psi_0$ obtains the mass  while all the other $SU(n)$ 
non-singlet components remain massless, so that 
the pair condensation of the $SO(2n=4\ell+2)$ chiral 
spinor fermion $\psi_\pm$ cannot 
realize the global minimum of the potential for $\ell\geq1$.

Also for the cases of $SO(2n)$ with even $n=2\ell$, it is easy to 
see that the $SU(n)$-invariant mass operator $M_n$ (\ref{eq:R0mass}) 
in the channel $\bfR_0$ can give a mass only to the $SU(n)$ singlet 
components and leaves the other non-singlet components 
(existing for $\ell\geq2$) massless. For this even 
$n=2\ell$ cases, however, the symmetric tensor product of $SO(2n)$ 
chiral spinors 
also contains other channels $\bfR_p$ ($p\geq1$) that contains
$SU(n)$-singlets.  
Since the only invariant tensors of $SU(n)$ is $\delta_i^j$ other than 
$\varepsilon^{i_1i_2\cdots i_n}$ and $\varepsilon_{i_1i_2\cdots i_n}$,   
the only possible $SU(n)$-invariant mass operators of lower rank than 
$a^1a^2\cdots a^n=\varepsilon^{i_1i_2\cdots i_n}a^{i_1}a^{i_2}\cdots 
a^{i_n}/n!$ 
must be constructed with the total number operator
$\hatN=\delta^i_ja^\dagger_ia^j$  
(rank 2). So the $SU(n)$-invariant mass operator contained 
in the channel $\bfR_p$ (rank $n-4p$ tensor) must be a polynomial 
$m( \hatN^q+c_1\hatN^{q-1}+\cdots+c_q)$ of degree $q=(n-4p)/2$. %
(See Appendix~\ref{Sec:SUn-singlet-in-tensor} for the explicit 
form of this $SU(n)$-singlet operators.)
In any case, those mass operators have eigenvalues depending on the 
$SU(n)$ fermion number so the $SU(n)$ multiplets have non-degenerate
masses so that the spinor pair condensation in those channels $\bfR_p$
($p\geq1$) cannot realize the global minimum of the potential.  

Although this general discussion is enough, let us demonstrate this 
general feature in an explicit example of $n=10$ case in which 
two bispinor channels $\bfR_1$ and $\bfR_2$ containing $SU(n)$-singlets 
exist other than the usual $\bfR_0$ of rank $n$. 

For $n=10$, i.e.,
$SO(20)\to SU(10)\times U(1)$, the $SO(20)$ irreducible spinors
are represented as follows in terms of $SU(10)$ fermion operators giving 
the branching rule:
\begin{equation}
\begin{array}{ccccccc}
\ket{\psi_+}&
\hspace{-.5em}=\,\ket0 \psi_0 
&\hspace{-.8em}+\, \displaystyle\frac12 a^\dagger_ia_j^\dagger\ket0 \psi^{ij}
&\hspace{-.8em}+\, (4\ a^\dagger\text{term})
&\hspace{-.8em}+\, (6\ a^\dagger\text{term})
&\hspace{-.8em}+\, (8\ a^\dagger\text{term})
&\hspace{-.8em}+\, a_1^\dagger a_2^\dagger\cdots a_9^\dagger a_{10}^\dagger\ket0 \tilde\psi_0 \\[1ex]
{\bf 512'}&\hspace{-.3em}=\,({\bf1})(-5) 
&\hspace{-1.5em}\oplus\ \ ({\bf45})(-3)
&\hspace{-.8em}\oplus\ \ ({\bf210})(-1)
&\hspace{-.5em}\oplus\ \, ({\bf\overline{210}})(1)
&\hspace{-.8em}\oplus\ \, ({\bf\overline{45}})(3)
&\hspace{-2.8em}\oplus\ \ \ \ \ ({\bf1})(5) \,, \\
\end{array}
\end{equation}
\begin{equation}
\begin{array}{ccccccc}
\ket{\psi_-}&
\hspace{-.5em}=\,a_i^\dagger\ket0 \psi^i  
&\hspace{-.8em}+\, \displaystyle\frac1{3!}a^\dagger_ia_j^\dagger a_k^\dagger\ket0 \psi^{ijk} 
&\hspace{-.8em}+\, (5\ a^\dagger\text{term})
&\hspace{-.8em}+\, (7\ a^\dagger\text{term})
&\hspace{-.8em}+\ \displaystyle
\frac1{9!}\varepsilon^{i_1i_2\cdots i_9j}a_{i_1}^\dagger a_{i_2}^\dagger\cdots a_{i_9}^\dagger 
\ket0 \tilde\psi_j & \\[1ex]
{\bf 512}&\hspace{-.8em}=\,({\bf 10})(-4) 
&\hspace{-1.5em}\oplus\ \ \ ({\bf120})(-2)
&\hspace{-1.5em}\oplus\ \ \ {\bf 252}(0)
&\hspace{-.5em}\oplus\ \  ({\bf\overline{120}})(2)
&\hspace{-2.8em}\oplus\ \ \ \ \ ({\bf\overline{10}})(4) \,.
& \\
\end{array}
\end{equation}
The symmetric tensor product of these $SO(20)$ spinors gives
\begin{align}
({\bf 512'}\otimes{\bf 512'})_S &= {\bf 92378'}(\bfR_0)
\oplus{\bf 38760}(\bfR_1)
\oplus {\bf190}(\bfR_2), \nn
({\bf 512}\otimes{\bf 512})_S &= {\bf 92378}(\bfR_0)\oplus{\bf 38760}(\bfR_1)
\oplus {\bf190}(\bfR_2). 
\end{align}
${\bf 92378'}(\bfR_0)$ (rank-10 tensor) contains the usual $SU(10)$
singlet which gives a mass operator
\begin{equation}
m_0\left(a^\dagger_1a^\dagger_2\cdots a^\dagger_{10}+ a^{10}a^9\cdots a^1\right) 
\end{equation}
for the spinor $\psi_+({\bf 512'})$, while ${\bf 92378}(\bfR_0)$ does not. 
The other channels $\bfR_1$ (rank-6 tensor) and $\bfR_2$ (rank-2 tensor) 
both contain $SU(10)$ singlets of the form (\ref{eq:Gamma6}) and
(\ref{eq:Gamma2}), respectively, which give the following mass operators  
for the both spinors $\psi_+({\bf 512'})$ and $\psi_-({\bf 512})$:
\begin{equation}
m_1 \left((\hatN-5)^3 -7(\hatN-5)\right)
+m_2 \left(\hatN-5\right),
\end{equation}
with $\hatN=\sum_{j=1}^5a^\dagger_ja^j$. 
Thus the 
mass matrix $M_+$ of the positive chiral fermion $\psi_+$ is given by
\begin{align}
&M_+ \ket{\psi_+}= 
\left[m_0\left(a^\dagger_1a^\dagger_2\cdots a^\dagger_{10}+ a^{10}a^9\cdots a^1\right) 
+m_1 \left((\hatN-5)^3 -7(\hatN-5)\right)
+m_2 \left(\hatN-5\right) \right]\ket{\psi_+}\nn
&\hspace{2em}=
\begin{pmatrix}
{-}90m_1{-}5m_2 & & & & & m_0 \\
& \hspace{-.8em}{-}6m_1{-}3m_2  & & & &     \\
&  & \hspace{-.8em}6m_1{-}m_2   & & &     \\
& & & \hspace{-.8em}{-}6m_1{+}m_2   & &     \\
& & & & \hspace{-.8em} 6m_1{+}3m_2  &     \\
m_0& & & & & \hspace{-.8em}90m_1{+}5m_2   \\
\end{pmatrix}
\begin{pmatrix}
\psi_0({\bf1})\\
\psi^{ij}({\bf45})\\
\psi^{(4)}({\bf210})\\
\psi^{(6)}({\bf\overline{210}})\\
\psi^{(8)}({\bf\overline{45}})\\
\tilde\psi_0({\bf1})\\
\end{pmatrix}.
\end{align}
The mass matrix $M_-$ of the negative chiral fermion $\psi_-$ is given by
\begin{align}
M_- \ket{\psi_-}&= 
\left[m_1 \left((\hatN-5)^3 -7(\hatN-5)\right)
+m_2 \left(\hatN-5\right)\right] \ket{\psi_-} \nn
&=
\begin{pmatrix}
{-}36m_1{-}4m_2 & & & & \\
& \hspace{-.8em}{-}6m_1{-}2m_2  & & & \\
&  & \hspace{-.8em}6m_1{-}m_2   & &   \\
& & & \hspace{-.8em}{-}6m_1{+}2m_2 &  \\
& & & & \hspace{-.8em} 36m_1{+}4m_2   \\
\end{pmatrix}
\begin{pmatrix}
\psi^i({\bf10})\\
\psi^{ijk}({\bf120})\\
\psi^{(5)}({\bf252})\\
\psi^{(7)}({\bf\overline{120}})\\
\tilde\psi_i({\bf\overline{10}})\\
\end{pmatrix}.
\end{align}
These masses proportional to 
$m_0,\ m_1$ and $m_2$ are generated by the fermion pair condensation 
in the bispinor channels $\bfR_0,\ \bfR_1$ and $\bfR_2$, respectively. 
Consider the case of pair condensation into a single channel. 
For the $\psi_+$ case, 
\begin{eqnarray}
\hbox{$\VEV{\bfR_0}\not=0$}&\rightarrow& \hbox{Only singlets $\psi_0({\bf1})$ and $\tilde\psi_0({\bf1})$ get non-zero mass square $m_0^2$} \nn  
&&\hbox{and other components ${\bf45}, {\bf210}, {\bf\overline{210}}, {\bf\overline{45}}$ remain massless}; \nn
\hbox{$\VEV{\bfR_1}\not=0$}&\rightarrow& \hbox{Singlets $\psi_0({\bf1})$ and $\tilde\psi_0({\bf1})$ get a mass square $(90m_1)^2$} \nn  
&&\hbox{and other components ${\bf45}, {\bf210}, {\bf\overline{210}}, {\bf\overline{45}}$ get another common mass$^2$ $(6m_1)^2$}; \nn
\hbox{$\VEV{\bfR_2}\not=0$}&\rightarrow& \hbox{All the $\psi_0({\bf1}), \tilde\psi_0({\bf1})$, 
${\bf45}, {\bf\overline{45}}$ and ${\bf210}, {\bf\overline{210}}$} \nn
&& \hbox{get different mass${^2}$ $(5m_2)^2,\ (3m_2)^2,\ (m_2)^2$, respectively}.
\end{eqnarray}
It is interesting that all the non-singlet components 
${\bf45}, {\bf\overline{45}}, {\bf210}, {\bf\overline{210}}$ 
have a degenerate mass for the $\bfR_1$ condensation. 
The situation is also similar for the $\psi_-$ case. 
In any case, however, the fermion mass spectrum does not realize the
total degeneracy, so the pair condensation in any channels $\bfR_0,
\bfR_1$ and $\bfR_2$ cannot realize the global minimum of the potential.

\section{Embedding-type subgroups}
\label{Sec:Embedding-type-subgroups}

Here we discuss ``embedding-type'' little group cases.
Since there are infinite ``embedding-type'' little groups,
we restrict this type up to $N=26$ for practicability.
Even for up to $N=26$, in view of 
Table~\ref{tab:SO-maximal-subgroups-simple-type}, we classify 
this ``embedding-type'' subgroups into seven sub-types; i.e., 
``orthogonal adjoint-type (rank-2 anti-symmetric tensor-type),''
``orthogonal rank-2 symmetric tensor-type,''
``symplectic rank-2 anti-symmetric tensor-type,''
``symplectic adjoint-type (rank-2 symmetric tensor-type),''
``unitary adjoint-type,'' 
``spinor-type,'' and 
``exceptional-type'' 
subgroups.

In the following, we summarize the branching rules of these 
sub-type subgroups. 
Several examples can be found in Ref.~\cite{Yamatsu:2015gut}.

\subsection{Orthogonal adjoint-type (rank-2 anti-symmetric tensor-type)}

We consider $SO\bigl(N=\frac{n(n-1)}{2}\bigr)\supset SO(n)$
$(n\in\mathbb{Z}_{\geq5})$ for $N\leq26$, 
in which the rank-2 anti-symmetric tensor representation
${\bf \frac{n(n-1)}2}$
of the subgroup $SO(n)$ is identified with the defining representation 
$\bfN$ of $SO(N)$. So the branching rule for the
$SO(N)$ vector $\bfN$ under this subgroup $SO(n)$ is
\begin{align}
\bfN={\bf \frac{n(n-1)}{2}}\,.
\end{align}
First, for the $n=5, N=10$ case, the breaking $SO(10)\to SO(5)$
does not occur via the pair condensation of the 
$SO(10)$ spinor fermion.
This is because the branching rule of 
the SO(10) (complex) spinor representation
${\bf16}$ or $\overline{\bf16}$ under 
$SO(5)(\simeq \USp(4))\subset SO(10)$ is
\begin{align}
 {\bf 16}={\bf (16)}, \qquad 
 {\bf \overline{16}}={\bf (16)}.
\end{align}
So the fermion pair $({\bf 16}\otimes{\bf 16})_{\rm S}$
of $SO(10)$ is viewed as 
a tensor product $({\bf 16}\otimes{\bf 16})_{\rm S}$ of $SO(5)$,  
but it has no $SO(5)$-singlet since 
$SO(5)$ ${\bf 16}$ is  a 
pseudo-real representation.

Second, consider the $n=6, N=15$ case. This time, the
$SO(15)\to SO(6)(\simeq SU(4))$ 
breaking can occur via the pair condensation of the $SO(15)$ 
spinor fermion {\bf128}  (real representation, now). 
This is because the branching rule of
$SO(15)\supset SO(6)(\simeq SU(4))$ for the spinor
representation is given by
\begin{align}
{\bf 128}=2{\bf (64)}.
\end{align}
$SO(15)$ spinor ${\bf 128}$ is real and so a tensor product
$({\bf 128}\otimes{\bf 128})_{\rm S}$ has a singlet of
$SO(15)$. $SO(6)$ ${\bf 64}$ is also a real representation, so
a tensor product 
$[({\bf 64}\oplus{\bf 64})\otimes({\bf 64}\oplus{\bf 64})]_{\rm S}$
of $SO(6)$ has three 
independent singlets of $SO(6)\subset SO(15)$. One of them is an
$SO(15)$ singlet.
From Table~\ref{Table:Representations-SOn-Spinor-Tensor}, 
the other $SO(6)$ singlets are in $\bfR_0={\bf 6435}$ and
$\bfR_1={\bf 455}$ representations, which  
correspond to rank-7 and rank-3 anti-symmetric tensor representations, 
respectively.
The fermion mass matrix from the rank-3 tensor representations $M_3$ is
traceless as noted before,  
so we have
\begin{align}
M_3\left(
 \begin{array}{c}
  |({\bf 64})\rangle_1\\
  |({\bf 64})\rangle_2\\
 \end{array}
 \right) 
 &\propto 
 \left(
 \begin{array}{c}
  +1\, |({\bf 64})\rangle_1\\
  -1\, |({\bf 64})\rangle_2\\
 \end{array}
 \right),
\end{align}
which yields the quadratic mass matrix $M_3^\dagger M_3$ proportional to
identity matrix. 
This can also be confirmed in another way as follows. 
Since $M_3$ is given by a linear combination of 
rank-3 gamma matrices 
$\Gamma_{a_1a_2a_3}$, 
the quadratic mass matrix 
$(M_3^\dag M_3)$ takes the form
\begin{align}
 M_3^\dag M_3=(\mbox{rank-6 part})+(\mbox{rank-4 part})
 +(\mbox{rank-2 part})+{\bf 1}\times(\mbox{\# of terms})
\end{align}
but the rank 6, 4, 2 parts must vanish since there are no such
$H$-singlets now, 
thus leaving only the part proportional to the identity matrix.
Also, the mass matrix from the rank-7 tensor representation $M_7$
is traceless, so the quadratic mass $(M_7^\dag M_7)$ is proportional to
the identity matrix.
Therefore, the VEVs of the auxiliary field $\Phi_{\bfR_0=\bf 6435}$ and
$\Phi_{\bfR_1=\bf 455}$ lead to
the identity quadratic mass matrix $(M_j^\dag M_j)\propto{\bf 1}$
$(j=3,7)$.

Third, consider the $n=7, N=21$ case. There is an 
$SO(21)\to SO(7)$
breaking via the pair condensation of the $SO(21)$ spinor fermion.
This is because the branching rule of
$SO(21)\supset SO(7)$ for the spinor representation is given by
\begin{align}
{\bf 1024}=2{\bf (512)}.
\end{align}
$SO(21)$ spinor ${\bf 1024}$ is pseudo-real, so a tensor product
$({\bf 128}\otimes{\bf 128})_{\rm S}$ has no singlet of
$SO(21)$. $SO(7)$ ${\bf 512}$ is a real representation, so
a tensor product
$[({\bf 512}\oplus{\bf 512})\otimes({\bf 512}\oplus{\bf 512})]_{\rm S}$
of $SO(6)$ has three independent singlets of $SO(7)\subset SO(21)$. 
From Table~\ref{Table:Representations-SOn-Spinor-Tensor}, the three
singlets of $SO(7)$ exist in $\bfR_0={\bf 352716}$,
$\bfR_1'={\bf 116280}$, and $\bfR_2'={\bf 1330}$ representations.
The VEVs of the scalar fields
$\Phi_{\bfR_0}$, $\Phi_{\bfR_1'}$, and $\Phi_{\bfR_2'}$ lead to
the quadratic mass matrix proportional to an identity,
$(M_j^\dag M_j)\propto{\bf 1}$
$(j=10,7,3)$, because of the traceless condition of $M_j$, again.

\subsection{Orthogonal rank-2 symmetric tensor-type}

We consider $SO\bigl(N=\frac{(n-1)(n+2)}{2}\bigr)\supset SO(n)$
$(n\in\mathbb{Z}_{\geq5\backslash6})$
for $N\leq26$, 
where the rank-2 symmetric tensor representation 
${\bf \frac{n(n+1)}2}$ of the subgroup $SO(n)$ is identified with the 
defining representation 
$\bfN$ of $SO(N)$. Then the branching rule for the
$SO(N)$ vector $\bfN$ under this subgroup $SO(n)$ is
\begin{align}
\bfN={\bf \frac{(n-1)(n+2)}{2}}.
\end{align}

First, for the $n=5, N=14$ case, there is no
$SO(14)\to SO(5)(\simeq \USp(4))$
breaking via the pair condensation of the $SO(14)$ spinor fermion.
This is because the branching rule of the $SO(14)$ (complex) 
spinor representation ${\bf64}$ or $\overline{\bf64}$ under 
$SO(5)(\simeq \USp(4))\subset SO(14)$ is
\begin{align}
{\bf 64}={\bf (64)}, \qquad 
{\bf \overline{64}}={\bf (64)}.
\end{align}
But the $SO(5)$ ${\bf 64}$ is a pseudo-real representation, so
a tensor product $({\bf 64}\otimes{\bf 64})_{\rm S}$ of $SO(5)$ has no
singlet of $SO(5)\subset SO(14)$.

Second, for the $n=6, N=20$ case, the breaking 
$SO(20)\to SO(6) (\simeq SU(4))$ can occur
via the pair condensation of the $SO(20)$ spinor fermion. 
This is because the branching rule of 
$SO(20)$ (pseudo-real) spinor ${\bf512}$ 
or ${\bf512}'$ under $SO(6)\subset SO(20)$ is given by
\begin{align}
{\bf 512}={\bf (256)}\oplus{\bf (\overline{256})},\qquad 
{\bf 512'}={\bf (256)}\oplus{\bf (\overline{256})}.
\end{align}
But $SO(6)$ ${\bf 256}$ and ${\bf \overline{256}}$
are complex representations, 
and the tensor product
$[({\bf 256}\oplus{\bf \overline{256}})
\otimes({\bf 256}\oplus{\bf \overline{256}})]_{\rm S}$ of
$SO(6)$ has one independent $SO(6)$ singlet, which is unique invariant 
${\bf \overline{256}}\cdot{\bf 256}$. 
From Table~\ref{Table:Representations-SOn-Spinor-Tensor}, the
singlet of $SO(6)$ exists in ${\bfR_1=\bf 38760}$ (rank-6 tensor) 
representation. 
The VEV of the auxiliary field 
$\Phi_{\bfR_1}$ must lead to
an identity-proportional quadratic 
mass matrix $(M_6^\dag M_6)\propto{\bf 1}$
since it gives the unique $SO(6)$ invariant mass 
${\bf \overline{256}}\cdot{\bf 256}$.

\subsection{Symplectic rank-2 anti-symmetric tensor-type}

We consider $SO\bigl(N=(n{-}1)(2n{+}1)\bigr)\supset\USp(2n)$
$(n\in\mathbb{Z}_{\geq3})$
for $N\leq26$, 
in which the rank-2 anti-symmetric $\Omega$-traceless tensor 
representation of the subgroup $\USp(2n)$,  
${\bf (n-1)(2n+1)}=\frac{2n(2n-1)}2-1$,  
is identified with the 
defining representation 
$\bfN$ of $SO(N)$. So the branching rule for the
$SO(N)$ vector $\bfN$ under this subgroup $\USp(n)$ is
\begin{align}
\bfN={\bf (n-1)(2n+1)}.
\end{align}
For the $n=3, N=14$ case, there is no
$SO(14)\to \USp(6)$
breaking via the pair condensation of the $SO(14)$ spinor fermion.
This is because the branching rule of $SO(14)$ (complex) 
spinor representation ${\bf64}$ or ${\bf \overline{64}}$
under $\USp(6)\subset SO(14)$ 
is given by
\begin{align}
{\bf 64}={\bf (64)},\qquad 
{\bf \overline{64}}={\bf (64)}.
\end{align}
But $\USp(6)$ ${\bf 64}$ is a pseudo-real representation, so
a tensor product $({\bf 64}\otimes{\bf 64})_{\rm S}$ of $\USp(6)$ has no 
$\USp(6)$ singlet.

\subsection{Symplectic adjoint-type (rank-2 symmetric tensor-type)}

We consider $SO\bigl(N=n(2n{+}1)\bigr)\supset\USp(2n)$
for $N\leq26$, 
in which the adjoint (rank-2 symmetric tensor)  
representation of the subgroup $\USp(2n)$,  
${\bf n(2n+1)}$,  
is identified with the 
defining representation 
$\bfN$ of $SO(N)$. So the branching rule for the
$SO(N)$ vector $\bfN$ under this subgroup $\USp(n)$ is
\begin{align}
\bfN={\bf n(2n+1)}\,.
\end{align}

First, for the $n=2, N=10$ case, there is no
$SO(10)\to \USp(4)$
breaking via the pair condensation of the $SO(10)$ spinor fermion.
This is because the branching rule of
$SO(10)$ (complex) spinor {\bf16} or $\overline{\bf16}$ under 
$\USp(4)\subset SO(10)$ is given by
\begin{align}
{\bf 16}={\bf (16)},\qquad 
{\bf \overline{16}}={\bf (16)}.
\end{align}
But $\USp(4)$ ${\bf 16}$ is a pseudo-real representation, so
a tensor product $({\bf 16}\otimes{\bf 16})_{\rm S}$ of $\USp(4)$ has no
$\USp(4)$ singlet.

Second, for the $n=3, N=21$ case, there is an 
$SO(21)\to \USp(6)$ 
breaking via the pair condensation of the $SO(21)$ spinor fermion.
This is because the branching rule of
$SO(21)\supset\USp(6)$
for the spinor representation is given by
\begin{align}
{\bf 1024}=2{\bf (512)}.
\end{align}
$SO(21)$ ${\bf 1024}$ is a pseudo-real representation, so a tensor product
$({\bf 1024}\otimes{\bf 1024})_{\rm S}$ of $SO(21)$ has no singlet of
$SO(21)$. $\USp(6)$ ${\bf 512}$ is a real representation, so
a tensor product
$[({\bf 512}\oplus{\bf 512})\otimes({\bf 512}\oplus{\bf 512})]_{\rm S}$
of $\USp(6)$ has three independent singlets of $\USp(6)\subset SO(21)$.
From Table~\ref{Table:Representations-SOn-Spinor-Tensor}, the three
singles of $\USp(6)$ exist in $\bfR_0={\bf 352716}$,
$\bfR_1'={\bf 116280}$, and $\bfR_2'={\bf 1330}$ representations. 
The VEVs of the auxiliary fields $\Phi_{\bfR_0}$, 
$\Phi_{\bfR_1'}$, and $\Phi_{\bfR_2'}$ lead to
the identity-proportional quadratic 
mass matrix $(M_j^\dag M_j)\propto{\bf 1}$
$(j=10,7,3)$ because of the tracelessness condition for each $M_j$.

\subsection{Spinor-type}

We consider
$SO\bigl(N=2^{\left[\frac{n-1}{2}\right]}\bigr)\supset SO(n)$,
$(n=8\ell, 8\ell\pm1, \ell\in\mathbb{Z}_{\geq1}; n\geq9)$, 
where the real spinor representation ${\bf 2^{\left[\frac{n-1}{2}\right]}}$
of the subgroup $SO(n)$, existing for $n=8\ell, 8\ell\pm1$,  
is identified with the 
defining representation $\bfN$ of $SO(N)$. So the branching rule for the
$SO(N)$ vector $\bfN$ under this subgroup $SO(n)$ is
\begin{align}
\bfN={\bf 2^{\left[\frac{n-1}{2}\right]}}.
\end{align}
For $SO(16)\supset SO(9)$, there is an $SO(16)\to SO(9)$
breaking via the pair condensation of the $SO(16)$ spinor fermion.
This is because the branching rule of 
$SO(16)$ real spinor, ${\bf128}$ or ${\bf128}'$, under  
$SO(9)\subset SO(16)$ is given by
\begin{align}
{\bf 128}={\bf (44)}\oplus{\bf (84)},\qquad 
{\bf 128'}={\bf (128)}.
\end{align}
Since $SO(16)$ ${\bf 128}$ is real,  a tensor product
$({\bf 128}\otimes{\bf 128})_{\rm S}$
of $SO(16)$ has one $SO(16)$ singlet, and contains two $SO(9)$ singlets
which come from ${\bf44}\otimes{\bf44}$ and ${\bf84}\otimes{\bf84}$. 
The $SO(9)$ singlets are in the rank-0 and rank-8 tensor representations.
The rank-0 tensor representation, proportional to
identity, is an $SO(16)$ singlet. 
The quadratic mass matrix from the rank-8 tensor representation
$(M_8^\dag M_8)$ is not proportional to identity and gives 
two different eigenvalues for ${\bf44}$ and ${\bf84}$ 
because of their 
dimension difference and the traceless condition for $M_8$.

$SO(16)$ ${\bf 128'}$ is a real representation, so a tensor product
$({\bf 128'}\otimes{\bf 128'})_{\rm S}$
of $SO(16)$ has a singlet of $SO(16)$. 
$SO(9)$ ${\bf 128}$ ($\gamma$-traceless vector-spinor), 
on the other hand, is also a real representation, so the fermion pair 
$({\bf 128'}\otimes{\bf 128'})_{\rm S}$ of $SO(16)$ spinor 
contains an $SO(9)$-singlet, but 
it is also a singlet of $SO(16)$, so does not cause the breaking 
$SO(16)\ \rightarrow\ SO(9)$.

\subsection{Unitary adjoint-type}

We consider $SO\left(N=n^2-1\right)\supset SU(n)$
for $N\leq26$, 
where the adjoint representation 
of the $SU(n)$ subgroup is identified with the 
defining representation 
$\bfN$ of $SO(N)$. So the branching rule for the
$SO(N)$ vector $\bfN$ under this subgroup $SU(n)$ is
\begin{align}
\bfN={\bf n^2-1}\,.
\end{align}

First, for the $n=3, N=8$ case, there is no $SO(8)\to SU(3)$
breaking via the pair condensation of the $SO(8)$ spinor fermion.
This is because the branching rule of
$SO(8)$ real spinor, ${\bf8}_s$ or${\bf8}_s$, under 
$SU(3)\subset SO(8)$ is given by
\begin{align}
{\bf 8_s}={\bf (8)},\qquad 
{\bf 8_c}={\bf (8)}.
\end{align}
Since $SO(8)$ spinors ${\bf 8_{s}}$ and ${\bf 8_{c}}$ are real, 
a tensor product
$({\bf 8_{s}}\otimes{\bf 8_{s}})_{\rm S}$ (or 
$({\bf 8_{c}}\otimes{\bf 8_{c}})_{\rm S}$) 
has a singlet of $SO(8)$.
$SU(3)$ adjoint ${\bf 8}$ is also a real representation, so 
the tensor product 
$({\bf 8}\otimes{\bf 8})_{\rm S}$ has 
an $SU(3)$ singlet, but it is also the $SO(8)$ singlet, so does not cause 
the breaking $SO(8)\ \rightarrow\ SU(3)$.

Second, consider the $n=4, N=15$ case. There is an 
$SO(15)\to SU(4)$ breaking via the pair condensation of the 
$SO(15)$ spinor fermion. 
This is because the branching rule of
$SO(15)$ real spinor representation {\bf 128} under   
$SU(4)\subset SO(15)$ is given by
\begin{align}
{\bf 128}=2{\bf (64)}.
\end{align}
$SO(15)$ spinor ${\bf 128}$ is real, so a tensor product
$({\bf 128}\otimes{\bf 128})_{\rm S}$ has an $SO(15)$ singlet. 
$SO(6)$ ${\bf 64}$ is also a real representation, so 
a tensor product 
$[({\bf 64}\oplus{\bf 64})\otimes({\bf 64}\oplus{\bf64})]_{\rm S}$
contains three independent $SU(4)$ singlets. One of them is the 
$SO(15)$ singlet.
From Table~\ref{Table:Representations-SOn-Spinor-Tensor}, the remaining
singlets of $SU(4)\simeq SO(6)$ exist in $\bfR_0={\bf 6435}$ and
$\bfR_1={\bf 455}$ representations. 
The VEVs of the scalar fields $\Phi_{\bfR_0}$ and
$\Phi_{\bfR_1}$ lead to
the quadratic mass matrix $(M_j^\dag M_j)\propto{\bf 1}$
$(j=7,3)$ proportional to identity because of
the traceless condition for $M_j$.

Third, for the $n=5, N=24$ case also, we quite similarly see that 
there is an $SO(24)\to SU(5)$ 
breaking via the pair condensation of the $SO(24)$ spinor fermion.
This is because the branching rule of
$SO(24)$ real spinor representation, {\bf2048} or ${\bf2048}'$, 
under $SU(5)\subset SO(24)$ is given by
\begin{align}
{\bf 2048}=2{\bf (1024)},\qquad 
{\bf 2048'}=2{\bf (1024)}.
\end{align}
$SO(24)$ spinor ${\bf 2048}$ is real, so a tensor product
$({\bf 2048}\otimes{\bf 2048})_{\rm S}$ has an $SO(24)$ singlet. 
$SU(5)$ ${\bf 1024}$ is also a real representation, so
a tensor product
$[({\bf 1024}\oplus{\bf 1024})\otimes({\bf 1024}\oplus{\bf 1024})]_{\rm S}$
has three independent $SU(5)$ singlets, one of which is 
the $SO(24)$ singlet.
From Table~\ref{Table:Representations-SOn-Spinor-Tensor}, the remaining
two $SU(5)$ singlets exist in $\bfR_0={\bf 1352078}^{(\prime)}$ and
$\bfR_1={\bf 735471}$ representations.
The VEVs of the scalar fields $\Phi_{\bfR_0}^{(\prime)}$,
and $\Phi_{\bfR_1}$ lead to
the quadratic mass matrix $(M_j^\dag M_j)\propto{\bf 1}$
$(j=12,7)$ proportional to identity because of 
the traceless property of $M_j$. 
Thus, their quadratic mass matrix is proportional to
identity and can realize the global minimum.

\subsection{Exceptional-type}

We consider exceptional-type maximal subgroups up to $N=26$. Only three
cases exist; 
$SO(7)\supset G_2$, $SO(14)\supset G_2$, and $SO(26)\supset F_4$.

First, consider $SO(7)\supset G_2$ case.  There is an
$SO(7)\to G_2$
breaking via the pair condensation of the $SO(7)$ spinor fermion. 
This is because the branching rule of
$SO(7)$ real spinor representation {\bf8} under  
$G_2\subset SO(7)$ is given by
\begin{align}
{\bf 8}={\bf (7)}\oplus{\bf (1)}.
\end{align}
$G_2$ singlet exists in ${\bf35}=\bfR_0$ (rank-3 tensor) in the fermion
pair $({\bf8}\otimes{\bf8})_S$ of $SO(7)$ spinor. 
 
We can apply the traceless condition to 
the mass matrix from the rank-3 
tensor representation $M_3$ also here and obtain
\begin{align}
M_3\left(
 \begin{array}{c}
  |({\bf 7})\rangle\\
  |({\bf 1})\rangle\\
 \end{array}
 \right) 
 &\propto 
 \left(
 \begin{array}{c}
  +1\, |({\bf 7})\rangle\\
  -7\, |({\bf 1})\rangle\\
 \end{array}
 \right).
\end{align}
Thus, the quadratic mass matrix $(M_3^\dag M_3)$ is not proportional to
identity, so
for this case, the pair condensation of the $SO(7)$ spinor cannot
realize the global minimum of the potential.

Second, consider the case $SO(14)\supset G_2$.  there is
an $SO(14)\to G_2$ breaking via the pair condensation of the 
$SO(14)$ spinor fermion. 
This is because the branching rule of
$SO(14)$ complex spinor representation, 
{\bf64} or $\overline{\bf64}$, 
under $G_2\subset SO(14)$ is given by
\begin{align}
{\bf 64}={\bf (64)},\qquad 
{\bf \overline{64}}={\bf (64)}.
\end{align}
$SO(14)$ spinor ${\bf 64}$ (or ${\bf \overline{64}}$) is complex,
 so a tensor product 
$({\bf 64}\otimes{\bf 64})_{\rm S}$ has no $SO(14)$ singlet,
while $G_2$ ${\bf 64}$ is a real representation, so
a tensor product $({\bf 64}\otimes{\bf 64})_{\rm S}$ of $G_2$ has one 
$G_2$ singlet. 
From Table~\ref{Table:Representations-SOn-Spinor-Tensor}, this $G_2$ singlet 
exists in $\bfR_1={\bf 364}$ (rank-3 tensor) representation. 
The VEVs of the scalar field $\Phi_{\bfR_1}$ 
must give the quadratic mass matrix
$(M_3^\dag M_3)\propto{\bf 1}$ proportional to identity since ${\bf64}$ is also 
irreducible under $G_2$. 
(Recall that the traceless condition does not apply to the chiral
 projected mass matrix $M_r$ for the $SO(N=8\ell\pm2)$ cases with
 complex spinor.)

Third, for $SO(26)\supset F_4$ case, the situation is 
almost the same as for 
the previous $SO(14)\supset G_2$ case.
There is an $SO(26)\to F_4$ 
breaking via the pair condensation of the $SO(26)$ spinor fermion.
This is because the branching rule of $SO(26)\supset F_4$
for the spinor representation is given by
\begin{align}
 {\bf 4096}={\bf (4096)},\qquad 
 {\bf \overline{4096}}={\bf (4096)}.
\end{align}
$SO(26)$ spinor ${\bf 4096}$ (or ${\bf \overline{4096}}$) is 
complex, so a tensor product 
$({\bf 4096}\otimes{\bf 4096})_{\rm S}$ of $SO(26)$ contains no $SO(26)$ 
singlet, while $F_4$ ${\bf 4096}$ is a real representation, so
a tensor product
$({\bf 4096}\otimes{\bf 4096})_{\rm S}$
of $F_4$ has one $F_4$ singlet. 
From Table~\ref{Table:Representations-SOn-Spinor-Tensor}, the 
$F_4$ singlet exists in 
$\bfR_1={\bf 3124550}$ (rank-9 tensor) representation. 
The VEV of the scalar field 
$\Phi_{\bfR_1}$ must give the quadratic mass matrix 
$(M_9^\dag M_9)\propto{\bf 1}$ proportional to identity since 
the $SO(26)$ spinor {\bf 4096} is also irreducible under $F_4$.

\section{Summary and discussions}
\label{Sec:Summary-discussion}

We summarize the symmetry breaking patterns in the $SO(N)$
NJL models whose fermions belong to an irreducible spinor representation
of $SO(N)$. 
The summary Table~\ref{Table:Summary-SOn-spinor-breaking} shows which
symmetry vacuum realizes (or vacua realize) the lowest potential energy
when which 4-Fermi coupling constant $G_{\bfR_k}$ beyond critical is the
strongest. 
For all of them, the spinor fermions get a totally degenerate mass so 
realizing the possible lowest potential energy.  
All the vacua are written there if they realize such a lowest potential 
energy in the $1/N$ leading order. The perturbation will determine the 
true vacuum among those degenerate vacua in the leading order.
From Table~\ref{Table:Summary-SOn-spinor-breaking},
we find that
when $SO(N)$ spinor representations are complex,
the $SO(N)$ symmetry is always broken into special subgroups.
When $SO(N)$ spinor representations are pseudo-real, 
the $SO(N)$ symmetry is broken into regular subgroups in 
most cases and into special subgroups in some cases.
When $SO(N)$ spinor representations are real, 
the $SO(N)$ symmetry is unbroken or broken into regular subgroups in
most cases and into special subgroups in some cases.
That is, our analysis have shown that the symmetry breaking into
special or regular subgroups is strongly correlated with the type of
representations; i.e., complex, or real, or
pseudo-real representations.

\section*{Acknowledgment}

This work was supported in part by the MEXT/JSPS KAKENHI Grant Number
JP18K03659 (T.K.), JP18H05543 and JP19K23440 (N.Y.).


{\small
\begin{center}
\begin{longtable}{lc|lcl}
 \caption{
 Maximal little groups $H$ or itself of $SO(N)$ spinor tensor
 product representations in the maximal subgroups of $SO(N) (N\leq26)$ 
 }
\label{Table:Representations-SOn-Spinor-Tensor}\\
\hline
\rowcolor[gray]{0.8}
$N$&R/PR/C&Maximal little groups $H$ of $SO(N)$&A/P/E&$H$-singlet in\\\hline
\endfirsthead
\multicolumn{5}{c}{Table~\ref{Table:Representations-SOn-Spinor-Tensor} (continued)}\\\hline
\rowcolor[gray]{0.8}
$N$&R/PR/C&Maximal little groups $H$ of $SO(N)$&A/P/E&$H$-singlet in\\\hline
\endhead
\hline
\endfoot
$3$&PR&
$SO(2)(R)$&A&${\bfR_0}$\\
\rowcolor[gray]{0.9}
$4$&PR&
$SU(2)\times U(1)(R)$&A&${\bfR_0}$\\
$5$&PR&
$SO(3)\times SO(2)(R)$&A&${\bfR_0}$\\
\rowcolor[gray]{0.9}
$6$&C&
 $SU(3)(R)$&A&${\bfR_0}$\\
\rowcolor[gray]{0.9}
&&$SO(3)\times SO(3)(S)$&P&${\bfR_0}$\\
$7$&R&
 $SO(4)\times SO(3)(R)$&A&${\bfR_0}$\\
&&$SO(7)(\text{No})$&A&${\bfR_1'}$\\
&&$G_2(S)$&E&${\bfR_0}$\\
\rowcolor[gray]{0.9}
$8$&R&
 $SU(4)\times U(1)(R)$&P&${\bfR_0}(*)$\\
\rowcolor[gray]{0.9} 
&&$SO(4)\times SO(4)(R)$&A&${\bfR_0}$\\
\rowcolor[gray]{0.9}
&&$SO(8)(\text{No})$&A&${\bfR_1'}$\\
$9$&R&
 $SO(5)\times SO(4)(R)$&A&${\bfR_0}$\\
&&$SO(8)(R)$&A&${\bfR_1'}$\\
&&$SO(9)(\text{No})$&A&${\bfR_1}$\\
&&$SO(3)\times SO(3)(S)$&P&${\bfR_0}$\\
&&$SU(2)(S)$&E&${\bfR_0}$\\
\rowcolor[gray]{0.9}
$10$&C&
 $SU(5)(R)$&P&${\bfR_0}$\\
\rowcolor[gray]{0.9}
&&$SO(5)\times SO(5)(S)$&A&${\bfR_0}$\\
\rowcolor[gray]{0.9}
&&$SO(9)(S)$&A&${\bfR_1}$\\
$11$&PR&
 $SO(6)\times SO(5)(R)$&A&${\bfR_0}$\\
&&$SO(9)\times SO(2)(R)$&A&${\bfR_1'}$\\
&&$SO(10)(R)$&A&${\bfR_1}$\\
\rowcolor[gray]{0.9}
$12$&PR&
 $SU(6)\times U(1)(R)$&P&${\bfR_0}(*)$, ${\bfR_1}$\\
\rowcolor[gray]{0.9}
&&$SO(6)\times SO(6)(R)$&A&${\bfR_0}$\\
\rowcolor[gray]{0.9}
&&$SO(10)\times SO(2)(R)$&A&${\bfR_1}$\\
$13$&PR&
 $SO(7)\times SO(6)(R)$&A&${\bfR_0}$\\
&&$SO(10)\times SO(3)(R)$&A&${\bfR_1'}$\\
&&$SO(11)\times SO(2)(R)$&A&${\bfR_1}$\\
\rowcolor[gray]{0.9}
$14$&C&
 $SU(7)(R)$&P&${\bfR_0}$\\
\rowcolor[gray]{0.9}
&&$SO(7)\times SO(7)(S)$&A&${\bfR_0}$\\
\rowcolor[gray]{0.9}
&&$SO(11)\times SO(3)(S)$&A&${\bfR_1}$\\
\rowcolor[gray]{0.9}
&&$G_2(S)$&E&${\bfR_1}$\\
$15$&R&
 $SO(8)\times SO(7)(R)$&A&${\bfR_0}$\\
&&$SO(11)\times SO(4)(R)$&A&${\bfR_1'}$\\
&&$SO(12)\times SO(3)(R)$&A&${\bfR_1}$\\
&&$SO(15)(\text{No})$&A&${\bfR_2'}$\\
&&$SO(5)\times SO(3)(S)$&P&${\bfR_0}$, ${\bfR_1'}$\\
&&$SO(6)(S)$&E&${\bfR_0}$, ${\bfR_1}$\\
&&$SU(2)(S)$&E&${\bfR_0}$, ${\bfR_1'}$, ${\bfR_1}$\\
\rowcolor[gray]{0.9}
$16$&R&
 $SU(8)\times U(1)(R)$&P&${\bfR_0}(*)$, ${\bfR_1}$\\
\rowcolor[gray]{0.9}
&&$SO(8)\times SO(8)(R)$&A&${\bfR_0}$\\
\rowcolor[gray]{0.9}
&&$SO(12)\times SO(4)(R)$&A&${\bfR_1}$\\
\rowcolor[gray]{0.9}
&&$SO(16)(\text{No})$&A&${\bfR_2}$\\
\rowcolor[gray]{0.9}
&&$SU(2)\times\USp(8)(S)$&P&${\bfR_0}(*)$, ${\bfR_1}$\\
\rowcolor[gray]{0.9}
&&$\USp(4)\times\USp(4)(S)$&P&${\bfR_0}(*)$, ${\bfR_1}$\\
\rowcolor[gray]{0.9}
&&$SO(9)(S)$&E&${\bfR_0}(*)$\\
$17$&R&
 $SO(9)\times SO(8)(R)$&A&${\bfR_0}$\\
&&$SO(12)\times SO(5)(R)$&A&${\bfR_1'}$\\
&&$SO(13)\times SO(4)(R)$&A&${\bfR_1}$\\
&&$SO(16)(R)$&A&${\bfR_2'}$\\
&&$SO(17)(\text{No})$&A&${\bfR_2}$\\
&&$SU(2)(S)$&E&${\bfR_0}$, ${\bfR_1'}$, ${\bfR_1}$\\
\rowcolor[gray]{0.9}
$18$&C&
 $SU(9)(R)$&P&${\bfR_0}$\\
\rowcolor[gray]{0.9}
&&$SO(9)\times SO(9)(S)$&A&${\bfR_0}$\\
\rowcolor[gray]{0.9}
&&$SO(13)\times SO(5)(S)$&A&${\bfR_1}$\\
\rowcolor[gray]{0.9}
&&$SO(17)(S)$&A&${\bfR_2}$\\
$19$&PR&
 $SO(10)\times SO(9)(R)$&A&${\bfR_0}$\\
&&$SO(13)\times SO(6)(R)$&A&${\bfR_1'}$\\
&&$SO(14)\times SO(5)(R)$&A&${\bfR_1}$\\
&&$SO(17)\times SO(2)(R)$&A&${\bfR_2'}$\\
&&$SO(18)(R)$&A&${\bfR_2}$\\
&&$SU(2)(S)$&E&${\bfR_0}$\\
\rowcolor[gray]{0.9}
$20$&PR&
 $SU(10)\times{U(1)}(R)$&P&${\bfR_0}(*)$, ${\bfR_1}$, ${\bfR_2}$\\
\rowcolor[gray]{0.9}
&&$SO(10)\times SO(10)(R)$&A&${\bfR_0}$\\
\rowcolor[gray]{0.9}
&&$SO(14)\times SO(6)(R)$&A&${\bfR_1}$\\
\rowcolor[gray]{0.9}
&&$SO(18)\times SO(2)(R)$&A&${\bfR_2}$\\
\rowcolor[gray]{0.9}
&&$SO(6)(S)$&E&${\bfR_1}$\\
$21$&PR&
 $SO(11)\times SO(10)(R)$&A&${\bfR_0}$\\
&&$SO(14)\times SO(7)(R)$&A&${\bfR_1'}$\\
&&$SO(15)\times SO(6)(R)$&A&${\bfR_1}$\\
&&$SO(18)\times SO(3)(R)$&A&${\bfR_2'}$\\
&&$SO(19)\times SO(2)(R)$&A&${\bfR_2}$\\
&&$SO(7)(S)$&E&${\bfR_0}$, ${\bfR_1'}$, ${\bfR_2'}$\\
&&$\USp(6)(S)$&E&${\bfR_0}$, ${\bfR_1'}$, ${\bfR_2'}$\\
&&$SU(2)(S)$&E&${\bfR_0}$, ${\bfR_1'}$, ${\bfR_1}$\\
\rowcolor[gray]{0.9}
$22$&C&
 $SU(11)(R)$&P&${\bfR_0}$\\
\rowcolor[gray]{0.9}
&&$SO(11)\times SO(11)(S)$&A&${\bfR_0}$\\
\rowcolor[gray]{0.9}
&&$SO(15)\times SO(7)(S)$&A&${\bfR_1}$\\
\rowcolor[gray]{0.9}
&&$SO(19)\times SO(3)(S)$&A&${\bfR_2}$\\
$23$&R&
 $SO(12)\times SO(11)(R)$&A&${\bfR_0}$\\
&&$SO(15)\times SO(8)(R)$&A&${\bfR_1'}$\\
&&$SO(16)\times SO(7)(R)$&A&${\bfR_1}$\\
&&$SO(19)\times SO(4)(R)$&A&${\bfR_2'}$\\
&&$SO(20)\times SO(3)(R)$&A&${\bfR_2}$\\
&&$SO(23)(\text{No})$&A&${\bfR_3'}$\\
&&$SU(2)(S)$&E&${\bfR_0}$, ${\bfR_1'}$, ${\bfR_1}$, ${\bfR_2'}$, ${\bfR_2}$\\
\rowcolor[gray]{0.9}
$24$&R&
 $SU(12)\times U(1)(R)$&P&${\bfR_0}(*)$, ${\bfR_1}$, ${\bfR_2}$\\
\rowcolor[gray]{0.9}
&&$SO(12)\times SO(12)(R)$&A&${\bfR_0}$\\
\rowcolor[gray]{0.9}
&&$SO(16)\times SO(8)(R)$&A&${\bfR_1}$\\
\rowcolor[gray]{0.9}
&&$SO(20)\times SO(4)(R)$&A&${\bfR_2}$\\
\rowcolor[gray]{0.9}
&&$SO(24)(\text{No})$&A&${\bfR_3}$\\
\rowcolor[gray]{0.9}
&&$\USp(6)\times\USp(4)(S)$&P&${\bfR_0}(*)$, ${\bfR_1}$, ${\bfR_2}$\\
\rowcolor[gray]{0.9}
&&$SO(3)\times SO(8)(S)$&P&${\bfR_0}$, ${\bfR_1}$, ${\bfR_2}$\\
\rowcolor[gray]{0.9}
&&$SU(2)\times\USp(12)(S)$&P&${\bfR_0}(*)$, ${\bfR_1}$, ${\bfR_2}$\\
\rowcolor[gray]{0.9}
&&$SU(5)(S)$&E&${\bfR_0}$, ${\bfR_1}$\\
$25$&R&
 $SO(13)\times SO(12)(R)$&A&${\bfR_0}$\\
&&$SO(16)\times SO(9)(R)$&A&${\bfR_1'}$\\
&&$SO(17)\times SO(8)(R)$&A&${\bfR_1}$\\
&&$SO(20)\times SO(5)(R)$&A&${\bfR_2'}$\\
&&$SO(21)\times SO(4)(R)$&A&${\bfR_2}$\\
&&$SO(24)(R)$&A&${\bfR_3'}$\\
&&$SO(25)(\text{No})$&A&${\bfR_3}$\\
&&$SO(5)\times SO(5)(S)$&P&${\bfR_0}$, ${\bfR_1'}$, ${\bfR_1}$, ${\bfR_2}$\\
&&$SU(2)(S)$&E&${\bfR_0}$, ${\bfR_1'}$, ${\bfR_1}$, ${\bfR_2'}$, ${\bfR_2}$\\
\rowcolor[gray]{0.9}
$26$&C&
 $SU(13)(R)$&P&${\bfR_0}$\\
\rowcolor[gray]{0.9}
&&$SO(13)\times SO(13)(S)$&A&${\bfR_0}$\\
\rowcolor[gray]{0.9}
&&$SO(17)\times SO(9)(S)$&A&${\bfR_1}$\\
\rowcolor[gray]{0.9}
&&$SO(21)\times SO(5)(S)$&A&${\bfR_2}$\\
\rowcolor[gray]{0.9}
&&$SO(25)(S)$&A&${\bfR_3}$\\
\rowcolor[gray]{0.9}
&&$F_4(S)$&E&${\bfR_1}$\\
\end{longtable}
\end{center}
{
Note that
R, PR and C in the R/PR/C column indicate that the 
$SO(N)$ spinor representation is real, pseudo-real and complex, 
respectively.
For $(*)$ attached subgroup $\bfR_0$ for $N=4k$ cases, which has two real 
(or pseudo-real) irreducible spinor $\bfr$ and $\bfr'$, only one
${\bfR}_0$  
either in $\bfr\times\bfr$ or in $\bfr'\times\bfr'$ contains the
$H$-singlet. 
A, P and E in the A/P/E column indicate addition-type, product-type, 
and embedding-type subgroup, respectively.
 $(R), (S)$ and (\text{No}) indicate that the subgroup $H$ is 
``regular'', ``special'' and ``No breaking'' (i.e.,$G=H$), respectively. 
 This list omitted the possible maximal little groups 
 which are not maximal subgroups, since those vacua in any case cannot 
 realize the  global minimum of the potential.
 }
}

\newpage

~
{\small
\begin{center}
\begin{longtable}{l|lllll}
\caption{Symmetry breaking pattern in 4D $SO(N)$ NJL model}
\label{Table:Summary-SOn-spinor-breaking}\\
\hline
\rowcolor[gray]{0.8}
{$N$}&
\hspace{3em}${\bfR_{0}}$&\hspace{4em}${\bfR_{1}^{(\prime)}}$&\hspace{4em}${\bfR_{2}^{(\prime)}}$&\hspace{3em}${\bfR_{3}^{(\prime)}}$\\\hline
\endfirsthead
\multicolumn{5}{c}{Table~\ref{Table:Summary-SOn-spinor-breaking} (continued)}\\\hline
\rowcolor[gray]{0.8}
{$N$}&
\hspace{3em}${\bfR_{0}}$&\hspace{4em}${\bfR_{1}^{(\prime)}}$&\hspace{4em}${\bfR_{2}^{(\prime)}}$&\hspace{3em}${\bfR_{3}^{(\prime)}}$\\\hline
\endhead
\hline
\endfoot
$3$&
$SO(2)(R)$\\
\rowcolor[gray]{0.9}
$4$&
$SU(2)\times U(1)(R)$\\
$5$&
$SO(3)\times SO(2)(R)$\\
\rowcolor[gray]{0.9}
$6$&
$SO(3)\times SO(3)(S)$\\
$7$&
$SO(4)\times SO(3)(R)$
&
${\bfR_{1}^{\prime}}:
 SO(7)(\text{No})
$\\
\rowcolor[gray]{0.9}
$8$&
$SO(4)\times SO(4)(R)$
&
$\phantom{\bfR_{1}:\,}SO(8)(\text{No})$
 \\
$9$&
$SO(5)\times SO(4)(R)$&
${\bfR_{1}^{\prime}}: SO(8)(R)$\\
&
$SO(3)\times SO(3)(S)$&
${\bfR_{1}}: SO(9)(\text{No})$\\
\rowcolor[gray]{0.9}
$10$&
$SO(5)\times SO(5)(S)$&
$\hspace{2.15em} SO(9)(S)$ \\
$11$&
$SO(6)\times SO(5)(R)$&
${\bfR_{1}^{\prime}}: SO(9)\times SO(2)(R)$ \\
&&
${\bfR_{1}}: SO(10)(R)$ \\
\rowcolor[gray]{0.9}
$12$&
$SO(6)\times SO(6)(R)$&
$\hspace{2.15em} SO(10)\times SO(2)(R)$
 \\
$13$&
$SO(7)\times SO(6)(R)$&
${\bfR_{1}^{\prime}}: SO(10)\times SO(3)(R)$ \\
&&
${\bfR_{1}}: SO(11)\times SO(2)(R)$ \\
\rowcolor[gray]{0.9}
$14$&
$SO(7)\times SO(7)(S)$&
$\hspace{2.15em} G_2(S)$ \\
\rowcolor[gray]{0.9}
&&
$\hspace{2.15em} SO(11)\times SO(3)(S)$ \\
$15$&
$SO(8)\times SO(7)(R)$&
${\bfR_{1}^{\prime}}: SO(11)\times SO(4)(R)$&
${\bfR_{2}^{\prime}}: SO(15)(\text{No})$\\
&
$SO(6)(S)$&
${\bfR_{1}}: SO(12)\times SO(3)(R)$&\\
&&
$\hspace{2.15em} SO(6)(S)$&\\
\rowcolor[gray]{0.9}
$16$&
$SO(8)\times SO(8)(R)$&
$\hspace{2.15em} SO(12)\times SO(4)(R)$&
$\hspace{2.15em} SO(16)(\text{No})$ \\
\rowcolor[gray]{0.9}
&&
$\hspace{2.2em} \USp(4)\times\USp(4)(S)$&\\
$17$&
$SO(9)\times SO(8)(R)$&
${\bfR_{1}^{\prime}}: SO(12)\times SO(5)(R)$&
${\bfR_{2}^{\prime}}: SO(16)(R)$ \\
&&
${\bfR_{1}}: SO(13)\times SO(4)(R)$&
${\bfR_{2}}: SO(17)(\text{No})$ \\
\rowcolor[gray]{0.9}
$18$&
$SO(9)\times SO(9)(S)$&
$\hspace{2.15em} SO(13)\times SO(5)(S)$&
$\hspace{2.15em} SO(17)(S)$ \\
$19$&
$SO(10)\times SO(9)(R)$&
${\bfR_{1}^{\prime}}: SO(13)\times SO(6)(R)$&
${\bfR_{2}^{\prime}}: SO(17)\times SO(2)(R)$ \\
&&
${\bfR_{1}}: SO(14)\times SO(5)(R)$&
${\bfR_{2}}: SO(18)(R)$ \\
\rowcolor[gray]{0.9}
$20$&
$SO(10)\times SO(10)(R)$&
$\hspace{2.15em} SO(14)\times SO(6)(R)$&
$\hspace{2.15em} SO(18)\times SO(2)(R)$ \\
\rowcolor[gray]{0.9}
&&
$\hspace{2.15em} SO(6)(S)$& \\
$21$&
$SO(11)\times SO(10)(R)$&
${\bfR_{1}^{\prime}}: SO(14)\times SO(7)(R)$&
${\bfR_{2}^{\prime}}: SO(18)\times SO(3)(R)$ \\
&
$SO(7)(S)$&
$\hspace{2.15em} SO(7)(S)$&
$\hspace{2.15em} SO(7)(S)$ \\
&
$\USp(6)(S)$&
$\hspace{2.15em} \USp(6)(S)$&
$\hspace{2.15em} \USp(6)(S)$ \\
&&
${\bfR_{1}}: SO(15)\times SO(6)(R)$&
${\bfR_{2}}: SO(19)\times SO(2)(R)$ \\
\rowcolor[gray]{0.9}
$22$&
$SO(11)\times SO(11)(S)$&
$\hspace{2.15em} SO(15)\times SO(7)(S)$&
$\hspace{2.15em} SO(19)\times SO(3)(S)$ \\
$23$&
$SO(12)\times SO(11)(R)$&
${\bfR_{1}^{\prime}}: SO(15)\times SO(8)(R)$&
${\bfR_{2}^{\prime}}: SO(19)\times SO(4)(R)$&
${\bfR_{3}^{\prime}}: SO(23)(\text{No})$ \\
&&
${\bfR_{1}}: SO(16)\times SO(7)(R)$&
${\bfR_{2}}: SO(20)\times SO(3)(R)$&\\
\rowcolor[gray]{0.9}
$24$&
$SO(12)\times SO(12)(R)$&
$\hspace{2.15em} SO(16)\times SO(8)(R)$&
$\hspace{2.15em} SO(20)\times SO(4)(R)$&
$\hspace{2.15em} SO(24)(\text{No})$ \\
\rowcolor[gray]{0.9}
&
$SU(5)(S)$&
$\hspace{2.15em} SU(5)(S)$&& \\
$25$&
$SO(13)\times SO(12)(R)$&
${\bfR_{1}^{\prime}}: SO(16)\times SO(9)(R)$&
${\bfR_{2}^{\prime}}: SO(20)\times SO(5)(R)$&
${\bfR_{3}^{\prime}}: SO(24)(R)$\\
&&
${\bfR_{1}}: SO(17)\times SO(8)(R)$&
${\bfR_{2}}: SO(21)\times SO(4)(R)$&
${\bfR_{3}}: SO(25)(\text{No})$ \\
\rowcolor[gray]{0.9}
$26$&
$SO(13)\times SO(13)(S)$&
$\hspace{2.15em} SO(17)\times SO(9)(S)$&
$\hspace{2.15em} SO(21)\times SO(5)(S)$&
$\hspace{2.15em} SO(25)(S)$ \\
\rowcolor[gray]{0.9}
&&
$\hspace{2em} F_4(S)$&& \\
\hline 
\end{longtable}
\end{center}
{In the row of indicated value $N$, the possible subgroups $H\subset G$ to which 
the symmetry $G=SO(N)$ is broken  
are shown in the column $\bfR_k^{(\prime)}$ for the cases where 
the coupling constant $G_{\bfR_k^{(\prime)}}$ 
is beyond critical and the strongest. 
$(R), (S)$ and (\text{No}) indicate that the subgroup $H$ is 
``regular'', ``special'' and ``No breaking'' (i.e.,$G=H$), respectively. 
$\bfR_k^{(\prime)}$ denotes both channels  
$\bfR_k$ and $\bfR'_k$, the latter channel $\bfR'_k$ with prime exists 
only for odd $N$. In the odd $N$ row, therefore, $\bfR_k$: and $\bfR'_k$: 
are put in front of the $H$ names to denote which channel of the two is 
the strongest. 
}

\appendix

\section{$SU(n)$-singlet operator in rank-$2k$ tensor $\Gamma_{a_1a_2\cdots a_{2k}}$ 
of $SO(2n)$}
\label{Sec:SUn-singlet-in-tensor}

$SU(n)$-singlet operator in rank-$2k$ tensor $\Gamma_{a_1a_2\cdots a_{2k}}$ 
can be found as follows. First, we see from 
$\Gamma_{2j-1}\Gamma_{2j}= i(2a_j^\dagger a^j-1)$ that the rank-2 $SU(n)$-singlet operator
is clearly given by
\begin{equation}
\bfGam^{(2)}:=\sum_{j=1}^n \Gamma^{(2)}_j 
= (2i)(\hatN - \frac{n}2), \qquad 
\Bigl( \Gamma^{(2)}_j :=\Gamma_{2j-1}\Gamma_{2j},\ \ 
\hatN := \sum_{j=1}^n a_j^\dagger a^j \Bigr).
\label{eq:Gamma2}
\end{equation} 
Then, it is also clear that the rank-$2k$ $SU(n)$-singlet operator 
for $k<n$ is given by
\begin{align}
\bfGam^{(2k)}&:=\sum_{1\leq j_1<j_2<\cdots<j_k\leq n}
\Gamma^{(2)}_{j_1}\Gamma^{(2)}_{j_2}\cdots \Gamma^{(2)}_{j_k}
=\frac1{k!}
\sum_{j_1,j_2,\cdots,j_k=1}^n 
\Gamma_{2j_1-1,{2j_1},{2j_2-1},{2j_2},\cdots,{2j_k-1},2j_k}\,.
\end{align}
Using the $\Gamma_a$ matrix algebra (or the fusion rule), we can derive
the identity
\begin{equation}
\bfGam^{(2)}\bfGam^{(2k)}=
(k+1)\bfGam^{(2k+2)}-(n-k+1)\bfGam^{(2k-2)}
\end{equation}
which recursively gives $\bfGam^{(2k)}$ as a polynomial of 
$\bfGam^{(2)}$ and hence as a polynomial of $(\hatN-n/2)$; e.g.,
\begin{align}
\bfGam^{(4)} &= \frac{(2i)^2}{2!}\left((\hatN-\frac{n}2)^2 
-\frac{n}4\right)\,, \\
\bfGam^{(6)} &= \frac{(2i)^3}{3!}\left((\hatN-\frac{n}2)^3 
-\frac{3n-2}4(\hatN-\frac{n}2)\right)\,, 
\label{eq:Gamma6}\\
\bfGam^{(8)} &= \frac{(2i)^4}{4!}\left((\hatN-\frac{n}2)^4 
-\frac{3n-4}2(\hatN-\frac{n}2)^2 
+\frac{3n(n-2)}{16}\right)\,. 
\end{align}
These expressions of $\bfGam^{(2k)}$ in terms of 
$\bfGam^{(2)}$ can also be obtained directly from the relation
\begin{equation}
\sum_{k=0}^n t^k \bfGam^{(2k)} = (1+t^2)^{\frac{n}2} e^{x\bfGam^{(2)}},
\label{eq:GeneratingFunc}
\end{equation} 
by expanding the RHS in powers in $t$, where $t=\tan{x}$ so that
\begin{equation}
x=\frac1{2i}\ln{\frac{1+it}{1-it}}
=t\sum_{\ell=0}^\infty\frac{(-t^2)^\ell}{2\ell+1}
= t-\frac{t^3}3+\frac{t^5}5+\cdots\,,\qquad \cos^2\!x= \frac{dx}{dt}=\frac1{1+t^2}.
\end{equation}
Noting that $\Gamma^{(2)}_j$ for each $j$ satisfies $(\Gamma^{(2)}_j)^2=-1$, 
we can obtain 
Eq.~(\ref{eq:GeneratingFunc}) as follows:
\begin{align}
e^{x\bfGam^{(2)}}
&=\prod_{j=1}^n e^{x\Gamma^{(2)}_j} 
=\prod_{j=1}^n \bigl(\cos x + \Gamma^{(2)}_j\sin x\bigr) 
=(\cos x)^n\prod_{j=1}^n \bigl( 1 + \Gamma^{(2)}_j\tan x\bigr) \nn
&=(\cos x)^n \sum_{k=0}^n \ (\tan x)^k 
\!\!\sum_{j_1<j_2<\cdots<j_k}\!\!
\Gamma^{(2)}_{j_1}\Gamma^{(2)}_{j_2}\cdots \Gamma^{(2)}_{j_k}
=(1+t^2)^{-\frac{n}2}\sum_{k=0}^n t^k\,\bfGam^{(2k)}.
\end{align}

\bibliographystyle{utphys} 
\bibliography{../../arxiv/reference}

\end{document}